\documentclass[11pt]{article}
\pdfoutput=1
\usepackage[T1]{fontenc}
\usepackage[utf8]{inputenc}
\usepackage{mathptmx}          
\usepackage[scaled=0.90]{helvet}
\usepackage{courier}
\usepackage{amsmath,amssymb,amsfonts}
\usepackage{graphicx}
\usepackage{booktabs}
\usepackage{multirow}
\usepackage{longtable}
\usepackage{array}
\usepackage{ragged2e}
\usepackage[table]{xcolor}
\usepackage[letterpaper,left=1.15in,right=1.15in,top=1.05in,bottom=1.05in]{geometry}
\usepackage[font=small,labelfont=bf,labelsep=colon,justification=justified]{caption}
\usepackage{fancyhdr}
\usepackage{titlesec}
\usepackage{placeins}
\usepackage{microtype}
\usepackage[super,comma,sort&compress]{natbib}
\usepackage[hidelinks,breaklinks=true]{hyperref}
\usepackage{url}

\fancypagestyle{plain}{%
  \fancyhf{}%
  \fancyfoot[C]{\footnotesize\thepage}%
}

\titleformat{\section}{\normalfont\large\bfseries}{\thesection}{0.6em}{}
\titleformat{\subsection}{\normalfont\normalsize\bfseries}{\thesubsection}{0.6em}{}
\titlespacing*{\section}{0pt}{1.6\baselineskip}{0.5\baselineskip}
\titlespacing*{\subsection}{0pt}{1.3\baselineskip}{0.4\baselineskip}
\titleformat{\paragraph}[runin]{\normalfont\normalsize\bfseries}{}{0pt}{}[.]
\titlespacing*{\paragraph}{0pt}{0.9\baselineskip}{0.5em}

\newcommand{\dg}{$^{\circ}$}
\newcommand{\bz}{\mathbf{z}}
\renewcommand{\arraystretch}{1.15}
\newcolumntype{L}[1]{>{\RaggedRight\arraybackslash}p{#1}}
\newcolumntype{C}[1]{>{\Centering\arraybackslash}p{#1}}

\newcommand{\papertitle}{Cross-species representation learning aligns mouse and human neural dynamics and tracks clinical drug efficacy}
\newcommand{\runninghead}{Cross-species representation learning tracks clinical drug efficacy}
\title{\bfseries\Large\papertitle}

\author{%
\normalsize Marko Tvrdic, Justin Richmond Domingo, Jae Ann Buenaluz, Jydell Ashley Palomo Penollar,\\[0.15em]
\normalsize Edmayelle Villavicencio Alforja, Jobi Fallaeria Subosa, Gabriel Ocana-Santero\textsuperscript{*}\\[0.6em]
\small\itshape Exin Therapeutics, Inc., 2261 Market Street STE 22614, San Francisco, California 94114, United States\\[0.35em]
\small\upshape\textsuperscript{*}Correspondence: \href{mailto:gabriel@exintherapeutics.com}{gabriel@exintherapeutics.com}%
}
\date{}

\hypersetup{pdftitle={Cross-species representation learning aligns mouse and human neural dynamics and tracks clinical drug efficacy},pdfauthor={Marko Tvrdic, Justin Richmond Domingo, Jae Ann Buenaluz, Jydell Ashley Palomo Penollar, Edmayelle Villavicencio Alforja, Jobi Fallaeria Subosa, Gabriel Ocana-Santero},pdfkeywords={cross-species representation learning; contrastive learning; EEG; translational biomarkers; epilepsy; drug efficacy},pdfsubject={Preprint}}

\begin{document}

\maketitle
\thispagestyle{plain}
\vspace{-1.2\baselineskip}

\begin{center}
\begin{minipage}{0.92\textwidth}
\section*{\centering\normalsize Abstract}
\vspace{-0.3\baselineskip}
\small
Preclinical models poorly predict human drug efficacy, particularly in neurological disorders. Neural activity offers a uniquely rich source of translational information because it captures high-dimensional variation in nervous-system function that can be measured in both animals and humans. However, its high dimensionality makes it difficult to distinguish conserved disease-related features from variation arising from species, recording modality and experimental context. Here, we test whether shared neural dynamics can be identified directly from electrophysiology data by learning representations organized by biological state rather than species. We develop a dual-rule contrastive learning framework that aligns corresponding mouse and human states while preserving separation between distinct phenotypes. This framework recovered conserved sensory-response structure across species and, in epilepsy, resolved distinct relationships between three mouse models and heterogeneous human patient populations. When treated animals were projected into a frozen cross-species representation, drug-induced movement towards the human-aligned healthy state retrospectively tracked known clinical efficacy across ten model-drug combinations including a disease-specific detrimental effect. The framework also identified shared disease-associated neural dynamics between \emph{Fmr1}-knockout mice and human 16p11.2 copy-number variant carriers despite differences in genetic aetiology and recording modality. Together, these findings show the potential of cross-species neural representation learning to map heterogeneous human disease onto experimentally tractable preclinical states and assess whether interventions restore human-relevant circuit function.
\end{minipage}
\end{center}

\vspace{0.8\baselineskip}

\section{Introduction}

Animal models are essential for studying disease mechanisms and evaluating therapeutic interventions before they can be studied in humans. Yet the ability of preclinical findings to predict human outcomes is limited. This is particularly evident in disorders of the central nervous system, where differences in brain organization, neural circuitry, behaviour and disease phenotype complicate translation across species\cite{r1,r2,r3,r4,r5,r6}. Even when analogous phenotypes can be measured in mice and humans, the features used to quantify them are often oversimplistic and selected independently in each species. This makes it difficult to determine whether an intervention corrects a conserved disease-relevant process or merely modifies a model-specific phenotype\cite{r7,r8}. Cross-species studies of electrophysiology have demonstrated that related neural processes can be measured in rodents and humans, but translating these measurements generally relies on a small number of predefined spectral, temporal or connectivity features\cite{r9,r10,r11,r12}.

Neural activity provides an attractive substrate for addressing this problem. Electrophysiological signals offer a direct, temporally resolved measure of circuit function, capturing dynamics that are difficult to resolve with molecular or behavioural readouts\cite{r13,r14}. However, correspondence at the level of measurement does not guarantee correspondence at the level of representation. Differences in brain geometry, electrode configuration, signal amplitude, behavioural context and recording environment create substantial species-specific variance in neural data. A sufficiently expressive machine-learning model can therefore obtain high predictive accuracy by exploiting features that distinguish experimental groups within one species while carrying little information about the corresponding biological state in another. For translational applications, the objective is thus different from conventional classification. Instead, the relevant representation should capture neural features related to the biological phenotype while ignoring differences that arise only from species or experimental conditions.

Representation learning offers a framework in which this requirement can be imposed directly on the geometry of the learned feature space\cite{r15,r16,r17,r18,r19}. Contrastive learning has proved effective for extracting informative representations from complex high-dimensional data by defining which observations should become similar or dissimilar during training, and supervised contrastive learning extends this principle by using known biological labels to organize the latent space\cite{r20,r21}. Related approaches have been used to align representations across domains, while contrastive objectives have also enabled common representations to be learned across participants and recording modalities in human neurophysiology\cite{r22,r23}.

For cross-species translation, however, the goal should not be to fully eliminate differences between species. Species identity is entangled with genuine biological differences, and forcing complete domain invariance may therefore remove signals relevant to the phenotype itself\cite{r24,r25}. A more useful goal is to align neural activity with constraints specifying precisely which structure should be invariant across species.

Here we introduce dual-rule contrastive neural representation learning, a framework for discovering cross-species biomarkers directly from neural activity. Our approach imposes two complementary learning objectives. First, neural recordings associated with the same biological state are brought together across species (e.g., healthy mouse - healthy human), encouraging the identification of features that are conserved between mouse and human. Second, recordings associated with different biological states are separated irrespective of species (e.g., healthy - disease), preserving the neural differences that distinguish health from disease. These two objectives organize neural activity into a shared representation in which corresponding biological states are aligned across species while distinct states remain separable.

We refer to representations satisfying these constraints as \emph{cross-species biomarkers}: latent features whose relationship to a biological state is conserved across species. This representation provides a common reference space in which the effects of therapeutics and other circuit perturbations can be evaluated across species. For example, an intervention administered only in mice can be projected into a latent space anchored by healthy and diseased human neural activity, allowing its effects to be quantified in relation to human disease. Rather than relying solely on improvements in conventional animal-model phenotypes, this approach allows us to assess whether an intervention shifts neural activity towards a state associated with human health. We hypothesized that such a measure of therapeutic response would be more predictive of human efficacy than conventional readouts.

We evaluate this framework across progressively more challenging translational settings. First, we collected matched human and mouse EEG during audiovisual sensory processing as a proof-of-concept, testing whether shared stimulus-related representations could be recovered despite substantial differences between recording domains. We then applied this approach to epilepsy, aligning multiple mouse models with human EEG to assess disease heterogeneity and whether pharmacological responses in the learned space could predict human efficacy and secondary effects. Finally, we tested generalization beyond closely matched disease models and recording modalities by aligning in-vivo electrophysiology recordings from \emph{Fmr1} knockout mice with scalp EEG recordings from individuals carrying 16p11.2 copy-number variants, asking whether convergent circuit dysfunction could be recovered despite distinct genetic aetiologies and measurements. Across these three settings we test whether cross-species representation learning can identify conserved neural dynamics, capture clinically-relevant disease heterogeneity, and relate drug effects in mice to clinical outcomes.

\section{Results}

\subsection{Dual-rule contrastive learning identifies shared latent neural dynamics of sensory perception across species}

\begin{figure}[t]
\centering
\includegraphics[width=\textwidth]{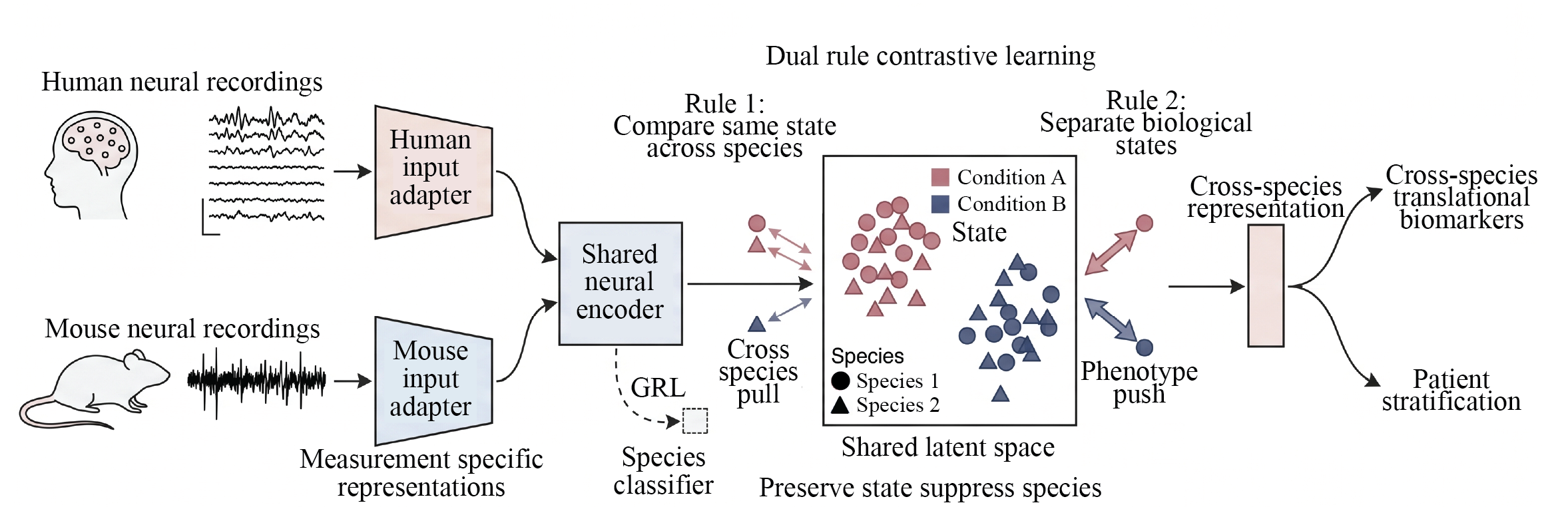}
\caption{\textbf{Dual-rule contrastive neural representation learning for cross-species biomarker discovery.} Schematic of the framework used to identify neural features that generalize between mice and humans.}
\label{fig:1}
\end{figure}

We first developed a representation-learning framework to identify neural features that capture shared biological variation across species, rather than features predictive only within mice or humans. Neural recordings from each species were mapped through measurement-specific input modules into a shared latent representation (Figure~\ref{fig:1}). We trained this representation using a dual-rule supervised contrastive objective that imposes two complementary constraints: alignment, whereby samples corresponding to the same biological state are preferentially brought together across species, and separation, whereby samples corresponding to different biological states are pushed apart irrespective of species. A gradient-reversal species classifier provided an additional regularization term discouraging species identity from dominating the shared representation. The resulting latent space is therefore optimized to preserve biological structure while suppressing species-specific variation. Once trained on matched biological states, this space can also serve as a reference into which unseen perturbations can be projected, providing a basis for assessing whether an experimental manipulation moves neural activity along a conserved, human-relevant axis (Figure~\ref{fig:1}).

\begin{figure}[p]
\centering
\includegraphics[width=0.77\textwidth]{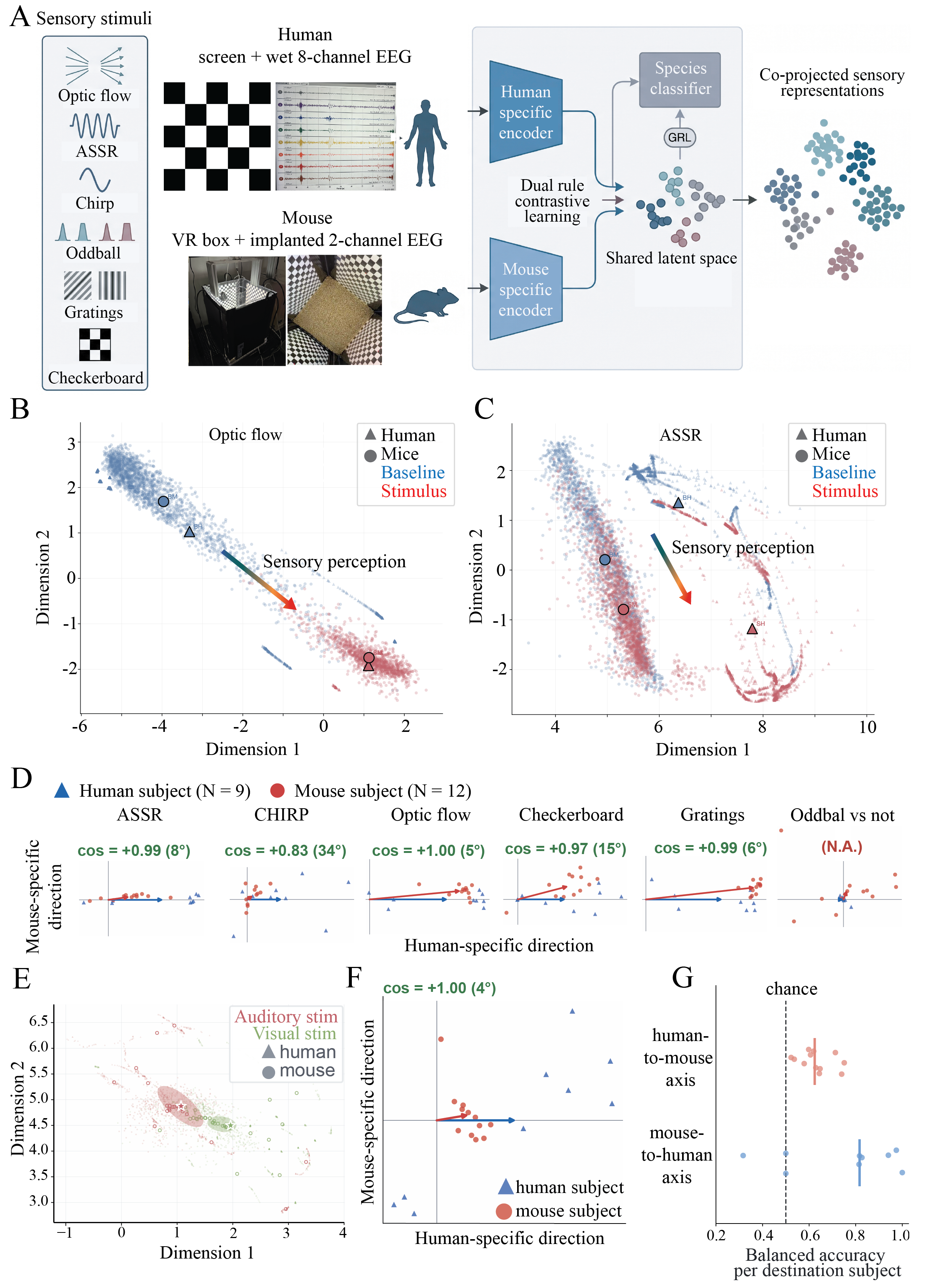}
\caption{\textbf{Dual-rule contrastive learning recovers conserved sensory-response axes across humans and mice} (A) Schematic of the experimental and computational framework. (B,C) Representative two-dimensional projections of the learned latent space showing baseline and stimulus-evoked activity in humans and mice for a visual paradigm (B, optic flow) and an auditory paradigm (C, ASSR). Large triangles and circles denote the corresponding human and mouse centroids. (D) Stimulus-versus-baseline response vectors for individual subjects across six sensory paradigms. Each point represents an individual subject projected onto the plane defined by the human- and mouse-derived mean response directions. Arrows indicate the species means. Cosine similarity and angular separation between the human and mouse mean directions are reported above each paradigm. (E) Joint embedding of auditory and visual stimulus responses in the coprojected mouse-human latent space. (F) Individual-subject projections onto the plane defined by the human- and mouse-derived mean sensory-response directions, pooled across sensory paradigms. Arrows indicate the species means. Cosine similarity and angular separation between the mean directions are shown above the plot. (G) Cross-species decoding of sensory state using an axis derived from one species and evaluated in held-out subjects from the other species. Points indicate balanced accuracy for individual destination subjects, and vertical bars indicate group summaries. The dashed line denotes chance performance (balanced accuracy = 0.5). Human $n$ = 9; mouse $n$ = 12.}
\label{fig:2}
\end{figure}

We initially tested whether such conserved axes could be recovered in a setting where the biological perturbation was experimentally matched across species. We collected EEG from healthy human volunteers and mice exposed to a common audiovisual stimulation paradigm (Figure~\ref{fig:2}A). Human participants ($n$ = 9) viewed stimuli on a screen while neural activity was recorded using an eight-channel wet EEG montage, whereas mice ($n$ = 12) were placed in a virtual-reality arena surrounded by visual displays and recorded using intracranial EEG. The paradigm comprised auditory steady-state responses (ASSR), frequency chirps and oddball stimuli, as well as optic flow, gratings and checkerboards. This design deliberately introduced substantial differences in species, recording geometry and experimental environment while preserving the underlying sensory manipulation, allowing us to ask whether the model could recover shared stimulus-related neural dynamics.

After training, baseline and stimulus-evoked recordings from the two species occupied a common representational space in which sensory stimulation produced directionally concordant shifts in humans and mice (Figure~\ref{fig:2}B,C). We quantified this correspondence by independently defining the mean stimulus-response direction in each species and measuring the cosine similarity between the resulting axes. Five of the six individual sensory paradigms showed strong cross-species directional alignment: ASSR, cosine similarity =0.99 (8\dg{} separation); chirp, 0.83 (34\dg); optic flow, 1.00 (5\dg); checkerboard, 0.97 (15\dg); and gratings, 0.99 (6\dg) (Figure~\ref{fig:2}D). In contrast, the oddball comparison did not yield a consistent cross-species axis, suggesting that the framework does not impose alignment when no consistent shared response is detected (Supplementary Table~\ref{tab:1}). We next attempted to enforce a similar separation modality-specific (i.e., audio vs visual). Indeed, when responses were aggregated across sensory paradigms, the human- and mouse-derived sensory axes were nearly collinear (cosine similarity =1.00, angular separation =4\dg; Figure~\ref{fig:2}E,F).

We next asked whether these shared directions generalized at the level of individual subjects rather than reflecting alignment of group centroids. For each paradigm, subjects from one species were projected onto an axis defined independently from the other species. Cross-species projections remained positive and significant after false-discovery-rate correction for ASSR, checkerboard, gratings and optic flow, as well as for the pooled auditory-versus-visual comparison (Supplementary Table~\ref{tab:1}). For example, mouse responses projected onto the independently-derived human axis with mean projections of approximately 0.64 for ASSR, 0.82 for checkerboards, 0.97 for gratings and 0.94 for optic flow; conversely, human responses projected positively onto the corresponding mouse-derived axes. Chirp responses showed weaker but directionally-consistent transfer, whereas oddball responses again failed to generalize across species. Consistent with these findings, sensory-response axes derived from one species also decoded sensory state above chance in held-out subjects from the other species (Figure~\ref{fig:2}G).

Dual-rule contrastive learning therefore identifies shared neural response dimensions across mice and humans despite differences in recording conditions. Having established this approach using matched sensory perturbations, we next asked whether it could identify clinically-relevant neural representations of disease and whether pharmacological interventions in mice could be evaluated in relation to human disease states.

\subsection{Cross-species neural biomarkers capture human-mouse disease state in epilepsy}

Having demonstrated that dual-rule contrastive learning can identify shared sensory-response dimensions across mice and humans, we next asked whether the same approach could capture disease-related neural activity. The translational value of a cross-species biomarker lies in its ability to predict human responses to therapeutic interventions from preclinical data. To achieve this, the biomarker must capture disease-related neural features shared between humans and experimental models. We investigated this in epilepsy, where abnormal neural activity is a defining feature of the disorder and EEG is central to clinical diagnosis\cite{r26}, providing a common measure of disease-related neural activity across species. Using the Temple University Hospital EEG datasets\cite{r27,r28} and recordings from three mechanistically-distinct mouse epilepsy models (PTZ, AY9944 and 4-AP)\cite{r29,r30,r31,r32}, we investigated how these mouse models captured different features of human epilepsy. These models were selected to capture different mechanisms and seizure types: PTZ induces a mixture of myoclonic, clonic, and tonic-clonic seizures through GABA$_{\mathrm{A}}$ receptor antagonism; AY9944 produces chronic atypical absence seizures characterized by slow spike-and-wave discharges; and 4-AP induces predominantly convulsive seizures, including clonic and tonic components, through blockade of voltage-gated potassium channels.

\begin{figure}[p]
\centering
\noindent\makebox[\textwidth][c]{\includegraphics[width=\textwidth]{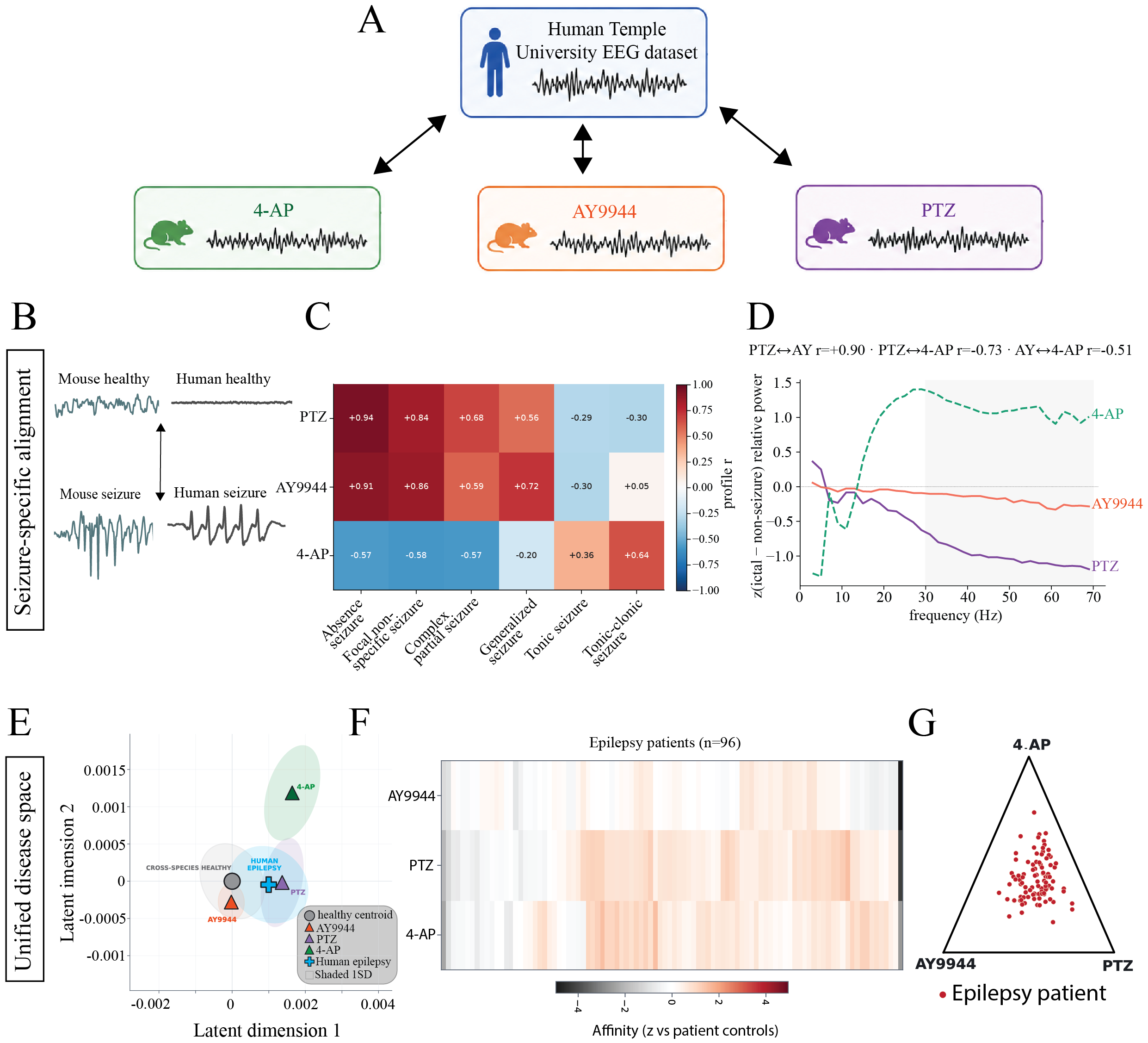}}
\caption{\textbf{Cross-species representation learning resolves conserved epilepsy signatures and model-specific patient heterogeneity.} A, Schematic of the cross-species framework, in which human EEG from TEMPLE/TUH is aligned independently with three mechanistically distinct mouse epilepsy models (AY9944, 4-AP and PTZ). B, Schematic of the seizure-specific alignment objective, jointly organizing mouse and human EEG ictal periods to compare shared features. C, Heatmap of cross-species associations between the three mouse epilepsy models and clinically annotated human seizure types. Values indicate standardized model-specific associations, with positive and negative values shown in red and blue, respectively. D, Frequency-dependent changes in spectral power during ictal relative to non-ictal activity in PTZ, AY9944, and 4-AP mice. Pairwise Pearson correlation coefficients between the model-specific spectral profiles are shown above the plot. E, Unified cross-species disease space showing the relative positions of healthy and epileptic human EEG and the three mouse models, illustrating a conserved disease direction together with model-specific structure. F, Heatmap of patient-level affinities to AY9944, PTZ, and 4-AP across 96 human epilepsy patients. Each column represents an individual patient, and values indicate affinity relative to patient controls. G, Ternary plot of the relative affinities of individual human epilepsy patients to the three mouse models. Each point represents one patient, with the three vertices corresponding to AY9944, PTZ, and 4-AP.}
\label{fig:3}
\end{figure}

We first examined whether the electrophysiological signatures of individual mouse epilepsy models corresponded to specific human seizure types. To do this, we compared ictal and non-ictal recordings from each mouse model with seizure-type-annotated human EEG from the Temple University Hospital Seizure Corpus (TUSZ). Recordings were processed using a common spectral frontend, with event durations matched across species to account for differences in seizure duration (Figure~\ref{fig:3}A,B). Each human seizure type was then evaluated separately against the ictal changes observed in each mouse model, allowing us to examine model-specific patterns of cross-species correspondence.

This analysis revealed distinct cross-species relationships. PTZ and AY9944 most strongly resembled human absence seizures, whereas 4-AP preferentially aligned with tonic-clonic and tonic seizures (Figure~\ref{fig:3}C). These relationships were mirrored by the underlying electrophysiology: PTZ and AY9944 produced similar frequency-dependent ictal changes, whereas the 4-AP signature was markedly different (PTZ--AY9944, $r$ = 0.90; PTZ--4-AP, $r$ = $-$0.73; AY9944--4-AP, $r$ = $-$0.51; Figure~\ref{fig:3}D). These findings indicate that different mouse models capture distinct aspects of human epileptic activity.

Although this analysis established correspondence with clinically defined seizure types, these labels capture only part of the heterogeneity of human epilepsy. We therefore sought to move beyond predefined seizure labels and compare human and mouse neural activity directly, incorporating both ictal and non-ictal recordings to identify disease-related variation.

We therefore next constructed a single cross-species latent space containing healthy and epileptic human EEG from the Temple University Hospital Epilepsy Corpus (TUEP) together with healthy and disease recordings from all three mouse models (Figure~\ref{fig:3}E). The learning objective imposed a hierarchy rather than forcing all pathological activity into a single cluster. Healthy mouse and human states were strongly co-aligned, disease states were encouraged to occupy a shared cross-species manifold, but the three mouse disease models were allowed to remain partially separated. A deliberately weak model-preference term then allowed each human epilepsy patient to lie close to one model, several models or a more broadly generic disease region. Species- and acquisition-specific information was assigned to a separate private latent space, while the shared representation was regularized to prevent collapse. This architecture was designed to preserve disease-related heterogeneity within the shared space, enabling patient-level comparisons rather than simply distinguishing epilepsy from healthy states.

The resulting geometry reflected that objective. PTZ and 4-AP disease states were robustly distinguishable from their matched controls in held-out animals (AUROC 0.994 and 1.000, respectively), whereas AY9944 separation was substantially weaker (AUROC 0.548). The latter is consistent with the weaker representation of the AY9944 disease signature in the human dataset, lacking examples of human patients with atypical absence seizures. Human epilepsy versus control classification was also deliberately modest (AUROC 0.623), consistent with the heterogeneity of a broad clinical epilepsy label and with an objective that preserves within-disease structure rather than optimizing binary classification. That retained structure enabled patient-level translation: across 96 individuals, affinities to AY9944, PTZ and 4-AP varied substantially, with some patients preferentially aligned to a single experimental model and others occupying intermediate positions (Figure~\ref{fig:3}F,G). These affinities were also associated with independent clinical features, including a positive relationship between PTZ affinity and antiseizure-medication burden and greater PTZ affinity among benzodiazepine-exposed patients (Supplementary Figure~\ref{fig:S1}). This is particularly interesting given that PTZ and benzodiazepines both target GABA-A receptors but with opposing effects\cite{rHansen}. Human epilepsy did not correspond to a single preclinical phenotype. Individual patients instead showed varying degrees of similarity to each of the three mouse models. This shared representation provides the basis for testing whether drug-induced shifts away from the mouse disease state and towards human healthy activity can predict therapeutic efficacy in patients.

\subsection{Cross-species representation learning tracks clinical drug efficacy from mouse neural dynamics in epilepsy models}

Having established a shared neural space linking individual patients to distinct preclinical disease states, we next asked the central translational question: does the effect of a drug on mouse neural dynamics in this cross-species space predict its efficacy in humans? To test this directly, we treated the three epilepsy models with a deliberately heterogeneous panel of antiseizure interventions spanning established clinical efficacy, seizure-type-specific activity and clinical failure\cite{r33} (Figure~\ref{fig:4}A). In 4-AP, we included two clinically-validated antiseizure medications with distinct mechanisms of action: valproate and ganaxolone\cite{r34}. We tested these alongside three investigational compounds with negative or disappointing clinical outcomes: JNJ-40411813 showed no significant benefit in a phase 2 study of focal epilepsy\cite{r35}, soticlestat missed its primary endpoints in phase 3 studies\cite{r36,r37}, and padsevonil failed to demonstrate sufficient efficacy in phase 2b/3 development\cite{r38}. In AY9944, whose neural phenotype shows a strong correspondence to human absence seizures, we selected compounds with contrasting clinical effects on absence epilepsy: valproate and ethosuximide, two established absence therapies\cite{r39} and tiagabine, which can aggravate absence seizures\cite{r40}. PTZ was treated with tiagabine and levetiracetam\cite{r41}, providing an additional mechanistically distinct test of pharmacological rescue. The panel was designed to determine whether the magnitude and direction of drug-induced changes in mouse neural dynamics could predict therapeutic efficacy in humans, including seizure-type-specific benefits and adverse effects.

All drug-response analyses were performed using a frozen cross-species representation. The model, preprocessing pipeline and disease manifolds were established exclusively from untreated recordings, with no retraining, fine-tuning or re-normalization following the introduction of treated EEG data. We then measured how treatment displaced each animal within the previously learned human-mouse geometry using two complementary approaches (Figure~\ref{fig:4}A-C). In AY9944 and PTZ, paired recordings before and after treatment enabled each animal's pretreatment state to be aligned with the disease manifold, allowing neural rescue to be quantified as the fraction of its individual disease-to-healthy trajectory traversed following drug administration. In 4-AP, where drugs were administered before seizure induction, the untreated disease state was unavailable for individual animals. Treatment effects were therefore quantified as displacement from the independently defined untreated 4-AP disease manifold, relative to vehicle-treated animals. In both approaches, the reference geometry was derived exclusively from untreated recordings, ensuring that drug-induced changes were evaluated as out-of-sample perturbations of the frozen cross-species representation.

The resulting neural shifts were pharmacologically coherent, and reproduced clinically important distinctions between compounds. In 4-AP, treatment groups differed significantly in their displacement from the disease state (Kruskal-Wallis $H$ = 12.21, $P$ = 0.032); valproate and ganaxolone produced significant rescue relative to vehicle ($P$ = 0.0268 and $P$ = 0.0255 respectively), whereas JNJ-40411813, soticlestat and padsevonil did not (Figure~\ref{fig:4}B). In the AY9944 and PTZ experiments, most therapeutically active conditions shifted individual animals toward healthy neural dynamics (Figure~\ref{fig:4}C). However, in the AY9944 absence-like model, tiagabine treatment displaced activity in the opposite direction from neural rescue. This is particularly informative because the propensity of tiagabine to exacerbate absence seizure activity was first described in preclinical models and subsequently reported in patients\cite{rGram,rCoenen,r40}, indicating that the cross-species representation recovers a clinically relevant pharmacological effect. In contrast, the same compound produced a positive rescue trajectory in PTZ. This model-dependent divergence shows that the cross-species representation captures the direction of a compound's therapeutic effect within a given disease state. Anticonvulsant activity alone does not resolve that direction, and resolving it is central to clinical translation.

We next examined whether the magnitude of preclinical neural rescue predicted the relative clinical efficacy of the tested interventions. Both the cross-species representation and the clinical efficacy benchmark were defined and frozen before any treated recordings were evaluated; treated data were never used to retrain, fine-tune or recalibrate the model, nor to revise the human efficacy labels. Because the paired AY9944/PTZ and 4-AP experiments required different rescue estimators, scores were standardized within the experimental family before pooling, and significance was assessed by permuting efficacy labels only within the family. Across all ten model-drug combinations, the degree of neural rescue strongly recapitulated the independently specified clinical efficacy ranking (Spearman $\rho$ = 0.87, blocked-permutation $P$ = 0.0039, bootstrap 95\% CI 0.43--0.90; Figure~\ref{fig:4}D). The same correlation was observed independently within both experimental families ($\rho$ = 0.87 in each). This ability to recover the relative ordering of clinically effective, ineffective and potentially detrimental interventions from their effects on mouse neural dynamics demonstrates the predictive potential of these metrics.

The cross-species biomarker showed a stronger association with human efficacy than conventional mouse readouts, including seizure count, high-amplitude time and time in seizure (Supplementary Figure~\ref{fig:S2}A-F). A biomarker extracted from the same latent space using mouse data alone performed poorly (Supplementary Figure~\ref{fig:S2}D), suggesting that mouse-specific features can obscure clinically relevant effects. Conventional readouts also tracked different drugs with varying success, highlighting the challenge of identifying which features of each mouse model translate to humans with traditional analysis. Agreement with human efficacy also increased with the number of animals per arm for the cross-species biomarker but not for the conventional readouts (Supplementary Figure~\ref{fig:S2}G), indicating that the biomarker may gain predictive power as sample sizes increase.

The geometry also suggested that treatment effects outside the intended rescue direction may contain complementary translational information. Displacement orthogonal to the disease-to-healthy trajectory (i.e., away from both healthy and disease centroids) showed a positive, although non-significant, association with an independently frozen measure of human off-target burden ($\rho$ = 0.54, $P$ = 0.10; Figure~\ref{fig:4}E). While exploratory, the direction of the effect is notable: movement along the conserved disease axis tracked therapeutic efficacy, whereas movement into dimensions not shared with the intended rescue trajectory tended to track secondary clinical effects (e.g., sedation). These results raise the possibility that cross-species neural biomarkers could provide a multidimensional pharmacological readout, simultaneously capturing therapeutic rescue and potentially identifying treatment-induced changes associated with adverse effects.

These results establish the translational potential of the cross-species framework. Without exposure to drug-response data, the frozen representation recovered the relative clinical efficacy of the tested interventions from treatment-induced changes in mouse neural dynamics, including a disease-specific adverse effect that could not be inferred from anticonvulsant activity alone. Cross-species biomarkers therefore offer a route from preclinical neural dynamics to human clinical neuropharmacology where therapeutic efficacy can be quantified through the correction of conserved disease features, while orthogonal perturbations may provide complementary information about clinically relevant secondary effects.

\clearpage
\phantomsection
\noindent\makebox[\textwidth][c]{\includegraphics[height=0.98\textheight]{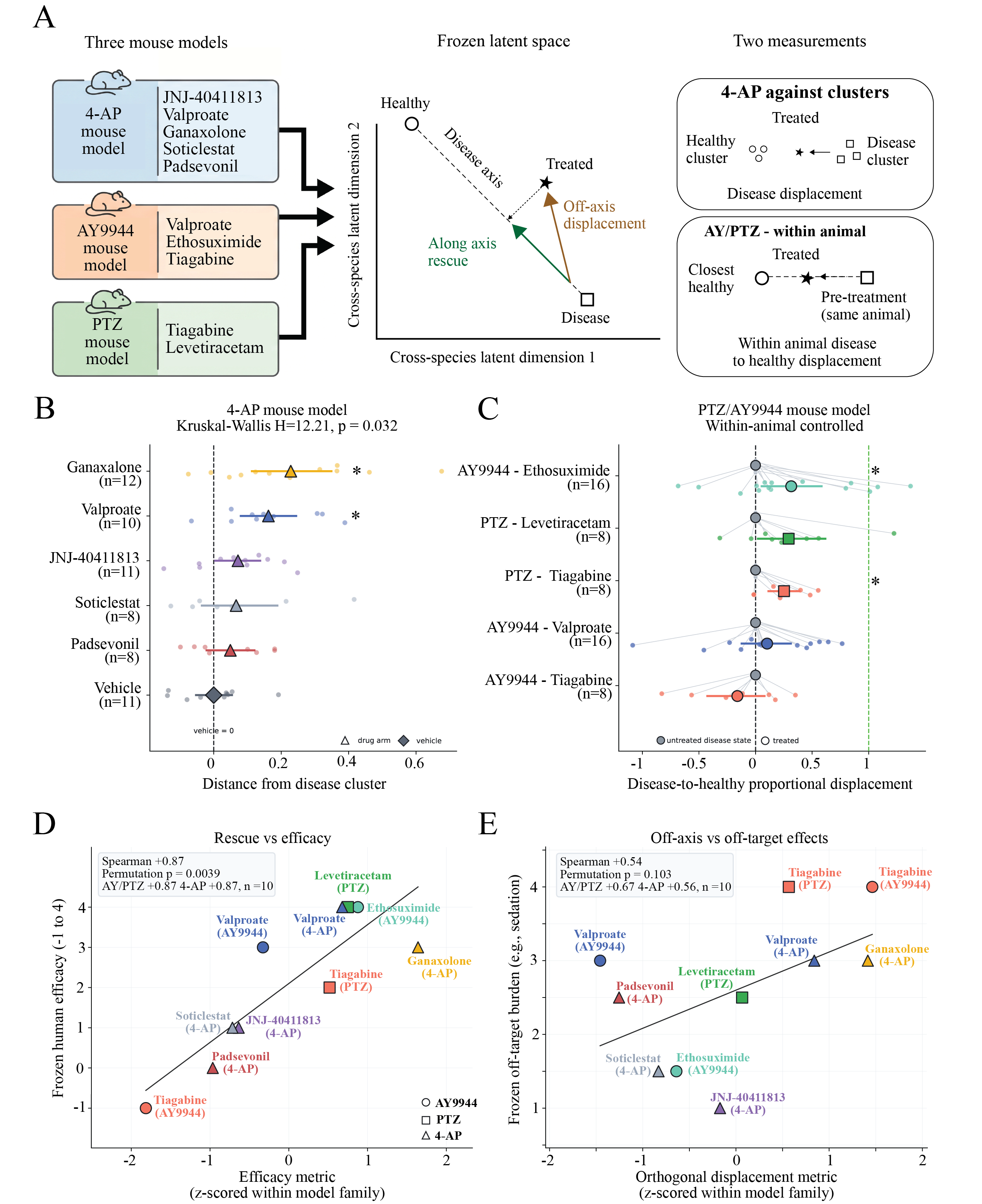}}\par
\vspace*{\fill}
\clearpage
\begingroup
\captionof{figure}{\textbf{Cross-species neural rescue in mice by antiseizure medications tracks clinical efficacy.} A, Schematic of the three epilepsy mouse models (4-AP, AY9944 and PTZ), the drugs tested in each model, and the quantification of treatment-induced displacement within a frozen cross-species latent space. Drug-induced movement is illustrated along the disease-to-healthy axis and orthogonal to this axis. The two measurement approaches are shown: displacement relative to the untreated disease cluster for 4-AP and within-animal disease-to-healthy displacement for AY9944 and PTZ. B, Distance from the untreated disease cluster relative to vehicle for the six treatment groups in the 4-AP model. Points represent individual animals, with larger symbols and horizontal bars indicating group estimates and uncertainty. Drug groups differed overall (Kruskal-Wallis $H$ = 12.21, $P$ = 0.032). Asterisks indicate significant differences from vehicle for valproate and ganaxolone. C, Within-animal disease-to-healthy proportional displacement following treatment in AY9944 and PTZ. Grey circles indicate pretreatment disease states, and coloured symbols indicate post-treatment positions. Individual animal measurements are connected by lines. The dashed vertical lines indicate the pretreatment disease state (0) and the closest healthy state (1). Asterisks indicate significant paired effects for ethosuximide in AY9944 and tiagabine in PTZ. D, Relationship between the preclinical neural rescue metric, standardized within experimental family, and the independently defined human efficacy score across ten model-drug combinations. Pooled Spearman $\rho$ = 0.87, blocked-permutation $P$ = 0.0039; within-family $\rho$ = 0.87 for both AY9944/PTZ and 4-AP. E, Relationship between the standardized orthogonal displacement metric and human off-target burden across the same ten model-drug combinations. Pooled Spearman $\rho$ = 0.54, blocked-permutation $P$ = 0.103; within-family $\rho$ = 0.67 for AY9944/PTZ and $\rho$ = 0.56 for 4-AP. In D and E, symbols distinguish the three mouse models, colours identify individual compounds, and solid lines indicate fitted linear trends.}
\label{fig:4}
\endgroup

\subsection{Cross-species representation learning generalizes across recording modalities and heterogeneous models of autism underlined by hyperexcitability}

\begin{figure}[tp]
\centering
\includegraphics[width=\textwidth]{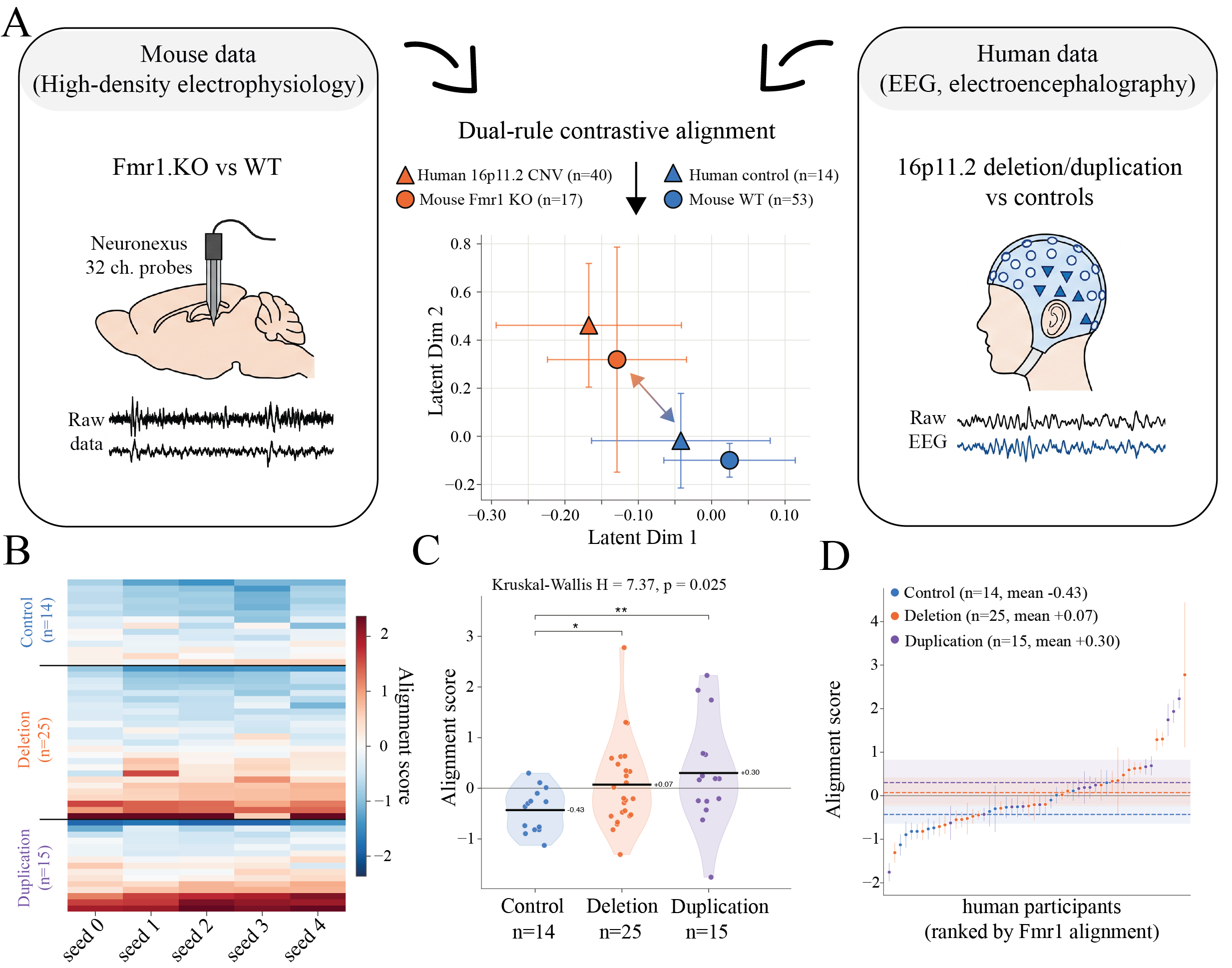}
\caption{\textbf{Cross-species alignment of \emph{Fmr1} mouse electrophysiology with human 16p11.2 EEG.} A, Schematic of the dual-rule contrastive learning framework aligning high-density electrophysiological recordings from \emph{Fmr1} knockout (KO) and wild-type (WT) mice with scalp EEG from individuals carrying 16p11.2 deletions or duplications and neurotypical controls. B, Participant-level alignment scores with the \emph{Fmr1}-associated disease representation across five independently trained models, grouped by human genotype. C, Distribution of mean alignment scores for neurotypical controls, 16p11.2 deletion carriers and duplication carriers. Alignment differed across groups (Kruskal-Wallis $H$ = 7.37, $P$ = 0.025). Points represent individual participants, and horizontal bars indicate group means. D, Individual participant alignment scores ranked by \emph{Fmr1}-associated alignment, grouped by human genotype. Dashed horizontal lines indicate group means (control, $-$0.43; deletion, +0.07; duplication, +0.30), with error bars indicating uncertainty in participant-level estimates.}
\label{fig:5}
\end{figure}

Having established that dual-rule contrastive learning could recover translationally meaningful disease structure in epilepsy, we next asked whether the framework could generalize across distinct recording modalities and disease aetiologies. We turned to two genetically distinct neurodevelopmental disorders, Fragile X syndrome, caused by loss of \emph{FMR1}, and 16p11.2 copy-number variation, a recurrent genomic alteration associated with autism and neurodevelopmental delay\cite{r42,r43,r44}. Despite their distinct genetic origins, both conditions have been associated with circuit hyperexcitability and altered excitation-inhibition balance\cite{r45,r46,r47,r48,r49}. We aligned high-density electrophysiological recordings from \emph{Fmr1} knockout mice with scalp EEG from individuals carrying 16p11.2 deletions or duplications and neurotypical controls (Figure~\ref{fig:5}A). This comparison deliberately crossed both recording modality and disease aetiology, testing whether the representation could identify conserved features of neural dysfunction across species, measurement geometries and distinct genetic perturbations.

Despite these differences, the learned representation identified a shared disease-associated direction. Across five independently trained models, control participants showed predominantly negative \emph{Fmr1}-associated alignment, whereas 16p11.2 deletion and duplication carriers shifted towards the mouse disease state (Figure~\ref{fig:5}B,C). Mean alignment scores were -0.43 in controls, +0.07 in deletion carriers and +0.30 in duplication carriers, with a significant overall difference between groups (Kruskal-Wallis $H$ = 7.37, $P$ = 0.025). Notably, alignment was strongest in duplication carriers, consistent with evidence that 16p11.2 duplication can produce impaired inhibitory transmission, cortical hyperexcitability and hypersynchronous network activity\cite{r50,r51}, features that converge with circuit abnormalities reported in Fragile X syndrome\cite{r45}.

Similarly to our epilepsy results, the representation also revealed substantial heterogeneity at the individual level. Ranking participants by their alignment with the \emph{Fmr1}-associated disease representation revealed broad, overlapping distributions within the deletion and duplication groups, with some carriers showing little correspondence to the mouse phenotype and others exhibiting strong alignment (Figure~\ref{fig:5}D). These findings suggest that the cross-species representation captures a continuous dimension of neural dysfunction that varies across individuals, rather than assigning a uniform disease signature to each genetic diagnosis.

These results extend the framework beyond epilepsy to genetically distinct neurodevelopmental disorders studied with different electrophysiological recording modalities. The cross-species representation recovered disease-associated structure in human scalp EEG from intracranial mouse recordings despite differences in species, measurement geometry and genetic aetiology. This supports the broader premise that shared features of neural dysfunction can be identified directly from electrophysiological dynamics, providing a potential route toward translational biomarkers that generalize across aetiologically-distinct disorders\cite{r52,r53}.

\FloatBarrier

\section{Discussion}

Here we show that cross-species neural representation learning can recover biologically-meaningful dimensions of neural activity that generalize between mice and humans despite substantial differences in species, recording modality and experimental context. Across increasingly challenging settings, the framework (1) recovered conserved sensory responses, (2) resolved distinct relationships between experimental epilepsy models and individual patients, and (3) identified convergent disease structure across genetically-distinct neurodevelopmental disorders. Most importantly, the frozen cross-species representation recovered the relative clinical efficacy of the tested interventions from drug-induced changes in mouse neural dynamics, without requiring any drug-response data for training or recalibration. These findings support the construction of a common biological coordinate system in which preclinical perturbations are evaluated by their relevance to human neural physiology.

This approach offers a novel perspective on the translational validity of animal models. Preclinical models are commonly evaluated according to how well they reproduce a disease as a whole\cite{r54}, yet heterogeneous neurological disorders are unlikely to correspond to a single experimental phenotype\cite{r55}. In epilepsy, PTZ, AY9944 and 4-AP occupied distinguishable regions of the learned disease space, and individual patients varied substantially in their affinity to each model. The relevant question may therefore be which patients, or which dimensions of their disease, a given model represents. With sufficiently diverse preclinical models, this could turn model selection into an empirical mapping problem in which experimental systems provide complementary coverage of human disease heterogeneity. The \emph{Fmr1}-16p11.2 analysis extends this idea further: translational correspondence need not require matched genetic aetiology, provided distinct perturbations converge on related circuit dynamics. Neural representations may therefore offer a means of organizing disease by functional pathophysiology rather than diagnosis or genotype alone\cite{r56}.

The same geometry provides a way to evaluate therapeutic efficacy. Conventional preclinical endpoints establish whether a treatment modifies an animal phenotype, but cannot determine whether the rescued feature is among those that are relevant to patients (Supplementary Figure~\ref{fig:S2}). Here, efficacy can instead be expressed as displacement towards a human-anchored healthy state. The opposite effects of tiagabine in PTZ and the absence-like AY9944 model illustrate why this distinction matters. The pharmacological effect of an intervention depends on the disease dynamics in which it acts. More broadly, the latent space describes treatment response in more than one dimension. Movement along the conserved disease-to-healthy trajectory tracked therapeutic efficacy, whereas orthogonal displacement showed a weaker relationship with clinical secondary effects. This second result is exploratory. If it holds, cross-species representations would separate therapeutic normalization from other pharmacological effects within a single coordinate system.

The next extension is prediction of therapeutic response for individual patients. Our results show that individual patients exhibit varying degrees of alignment with experimentally tractable disease states, while pharmacological interventions differ in their ability to restore healthy neural dynamics within those same states. Combining these two relationships suggests a potential route to personalized treatment prediction: matching individual patients to the preclinical disease states that most closely resemble their neural dynamics and identifying the interventions that most effectively restore healthy activity within those states\cite{r57,r58,r59}. In this formulation, therapeutic prediction becomes an out-of-sample, or effectively zero-shot, inference problem rather than conventional supervised prediction from large cohorts of treated patients. Such an approach could also support clinical-trial enrichment by identifying patients whose baseline neural state most closely resembles the preclinical state in which a therapy produces rescue\cite{r57,r58,r59}. The present study establishes the foundation for this approach. Prospective prediction of treatment response in individual patients is the next step.

Several limitations define the scope of these conclusions. The pharmacological analysis was restricted to ten model-drug combinations and relied on retrospective clinical efficacy estimates rather than prospectively-collected human responses. Furthermore, the clinical efficacy benchmark combined interventions evaluated across different epilepsy syndromes and clinical endpoints. The epilepsy representation showed modest human disease discrimination, and AY9944 was weakly separated from control in the unified space, emphasizing that cross-species alignment cannot recover biological information that is poorly represented in the source datasets. External validation across recording systems, additional disease models and substantially larger intervention panels will therefore be important. A further conceptual limitation is the assumption that successful therapy moves neural activity towards an average healthy state. Effective interventions may instead produce compensatory neural states that restore function without reproducing untreated healthy dynamics. Future approaches will then need richer representations of the healthy and disease manifolds rather than a single restorative axis. Nevertheless, these findings support a shift in the central question of translational pharmacology: from asking whether a drug rescues an animal phenotype to determining which components of its therapeutic effects are conserved across species and relevant to human clinical efficacy. Scaling this framework across diverse preclinical models, pharmacological perturbations and human populations could make preclinical drug development a patient-anchored prediction problem, linking experimentally tractable neural states to individualized therapeutic responses in humans.

\section{Methods}

\paragraph{Study design} We developed a cross-species representation-learning framework to identify neural features whose relationship to biological state is conserved between mice and humans. The framework was evaluated in three settings: matched audiovisual stimulation, epilepsy across three mouse models and human EEG, and a cross-aetiology neurodevelopmental comparison between \emph{Fmr1}-knockout mice and individuals carrying 16p11.2 copy-number variants.

Train, validation and test partitions were defined at the participant or animal level. For the unified epilepsy model, three fixed subject-level folds were generated once and reused across model seeds and configurations, so that seeds varied model initialization, minibatch order and sampled windows but not data partitions. For the \emph{Fmr1}--16p11.2 model, five subject-level folds were redrawn for each seed. No windows from a single subject were distributed across partitions.

\paragraph{Sensory EEG datasets} EEG was recorded from nine healthy human participants (6 male and 3 female) using an eight-channel OpenBCI wet-cap system at 250 Hz and from mice using two implanted EEG channels, with additional EMG and accelerometer signals acquired at 1,000 Hz. Mouse signals were resampled to 250 Hz. The reported analyses used EEG channels only.

The sensory battery comprised auditory steady-state responses (ASSR), chirps, auditory oddball stimuli, optic flow, gratings and checkerboards. Epoch durations were matched where possible across species: 2 s for ASSR, 2 s for oddball, 1 s for gratings, 1.5 s for optic flow and 20 s for checkerboards; chirps were 4 s in humans and 5 s in mice. Baseline recordings were 20 s and modality-specific in humans and 5 s from a common mouse baseline. Twenty-one mice had visual recordings, but baseline data required for the reported baseline-versus-stimulus analysis were available for 12 animals; analyses were therefore restricted to these 12 mice.

\paragraph{Epilepsy datasets and preprocessing} Human epilepsy EEG was obtained from the Temple University Hospital datasets. TUSZ was used for seizure-type-specific analyses, including absence (ABSZ, $n$ = 11), focal non-specific ($n$ = 52), complex partial ($n$ = 39), generalized non-specific ($n$ = 49), tonic ($n$ = 4) and tonic-clonic seizures ($n$ = 15). Ictal intervals were restricted to the target seizure type and interictal EEG was sampled at least 60 s from any annotated seizure. TUEP was used for the unified patient-level disease representation without seizure-type labels.

Mouse EEG was obtained from PTZ, 4-AP and AY9944 epilepsy models as in previous work. PTZ discharges were contrasted with duration-matched non-seizure activity from the same animals; 4-AP ictal activity was contrasted with within-animal non-seizure activity; and AY9944 spike-wave discharges were contrasted with identically detected events in saline controls.

Epilepsy recordings were resampled to 250 Hz, zero-phase band-pass filtered from 0.5-70 Hz using a fourth-order Butterworth filter and robustly normalized per channel using the median and interquartile range. No manual artefact rejection or hand-selection of clean EEG epochs was performed.

\paragraph{Dual-rule contrastive representation learning} Species-specific input modules mapped human and mouse measurements into a shared encoder producing latent representation $\bz$. Training followed two principles: observations representing the same biological state were encouraged to align across species, whereas biologically distinct states remained separated.

For anchor $i$, the general form of the contrastive loss was
\begin{equation}
\mathcal{L}_{i} \;=\; -\log
\frac{\displaystyle\sum_{p \in \mathcal{P}(i)} w_{ip}\,
\exp\!\left[\operatorname{sim}(\bz_{i},\bz_{p})/\tau\right]}
{\displaystyle\sum_{a \neq i}
\exp\!\left[\operatorname{sim}(\bz_{i},\bz_{a})/\tau\right]},
\label{eq:contrastive}
\end{equation}
where $\mathcal{P}(i)$ denotes observations sharing the biological label of $i$, $\tau$ is the contrastive temperature and cross-species positive pairs were upweighted by $w_{ip} = 1 + \beta\,\mathbf{1}(s_{i} \neq s_{p})$. The two applications instantiated this objective differently, as detailed in Supplementary Table~\ref{tab:3}.

The unified epilepsy model contained private representations for acquisition-specific information and anti-collapse constraints based on latent variance, covariance and participation ratio. Checkpoints were selected using validation subjects only; held-out patient-alignment and all drug-response metrics were excluded from model selection. Dataset composition, preprocessing, architecture, loss weights, optimization, cross-validation and evaluation settings for both the unified epilepsy model and the \emph{Fmr1}--16p11.2 model are given in Supplementary Table~\ref{tab:3}.

\paragraph{Representation performance} Balanced accuracy of the learned representations on the training units and on held-out subjects (95\% confidence intervals where estimated).

\begin{center}
\footnotesize
\begin{tabular}{lcc}
\toprule
\textbf{Contrast} & \textbf{Train} & \textbf{Held-out} \\
\midrule
\emph{Fmr1} human, CNV vs control & 0.80 & 0.59 [0.38, 0.78] \\
\emph{Fmr1} mouse, KO vs WT       & 0.90 & 0.80 [0.66, 0.92] \\
Epilepsy, human                   & 0.72 & 0.62 [0.54, 0.70] \\
Epilepsy, AY9944                  & 0.58 & 0.55 [0.41, 0.67] \\
Epilepsy, PTZ                     & 0.97 & 0.99 \\
Epilepsy, 4-AP                    & 0.95 & 1.00 \\
\bottomrule
\end{tabular}
\end{center}

\paragraph{Sensory cross-species alignment} The sensory model additionally used a gradient-reversal species classifier on the shared latent to suppress species identity (Figure~\ref{fig:1}). For each sensory paradigm, stimulus-response directions were estimated independently in humans and mice from the difference between stimulus and baseline latent representations. Cross-species agreement was quantified by cosine similarity between these directions. Subject-level generalization was assessed by projecting individuals from one species onto an axis estimated independently in the other species. Cross-species decoding similarly used an axis learned in one species to classify held-out subjects from the other.

\paragraph{Seizure-type-specific spectral correspondence} The seizure-type correspondence analysis in Figure~\ref{fig:3} was performed separately from the learned representation. Welch power spectral density was summarized into 34 contiguous 2-Hz bins spanning 2--70 Hz\cite{r60}. Power was expressed as log relative power and standardized by frequency bin.

Human seizures were divided into consecutive epochs whose durations were sampled from the corresponding mouse event-duration distribution; matched interictal segments were drawn from the same recording. Mouse disease-versus-reference and human seizure-versus-interictal contrasts produced 34-dimensional spectral difference profiles. Model--seizure correspondence was quantified as the Pearson correlation between these profiles. As a control for differences in aperiodic spectral structure, profiles were linearly detrended against log frequency before recomputing correlations.

Separately, dual-rule models were trained for the corresponding mouse--human contrasts and evaluated by bidirectional transfer AUROC. Thus, spectral-profile similarity and learned cross-species transfer constituted independent measures of correspondence.

\paragraph{Unified epilepsy representation and patient affinity} A unified representation was trained from human TUEP EEG together with healthy and disease activity from PTZ, 4-AP and AY9944 mice. Healthy activity was strongly aligned across species, whereas the three mouse disease states were permitted to retain model-specific structure.

EEG was represented using 8-s windows. Each analysis unit contributed equal-N samples ($N$ = 64), pooled using a DeepSets\cite{r61} architecture into a 14-dimensional translational representation. Patient and prospective treatment embeddings were calculated from 50 repeated equal-N window draws and averaged in latent space, preventing recording duration or window count from determining affinity.

Patient affinity quantified the position of each individual relative to the model-specific disease and reference representations. The geometry was estimated from training units $\mathcal{T}$ only, giving a shared healthy centroid $C_H$, per-model disease and control centroids $C_m^{D}$ and $C_m^{C}$, and the mean per-dimension standard deviation $s$ of the training disease units:
\begin{equation*}
\hat{v}_m=\frac{C_m^{D}-C_m^{C}}{\lVert C_m^{D}-C_m^{C}\rVert},
\qquad
\operatorname{dist}_m(\bz)=\frac{\lVert \bz-C_m^{D}\rVert}{s},
\qquad
\operatorname{proj}_m(\bz)=(\bz-C_H)\cdot\hat{v}_m .
\end{equation*}
Each quantity was z-scored against the training human controls (denoted by a tilde), so that $s$ cancels. Affinity and the soft model assignment were
\begin{equation*}
A_m(\bz)=\tfrac{1}{2}\big[\widetilde{\operatorname{proj}}_m(\bz)-\widetilde{\operatorname{dist}}_m(\bz)\big],
\qquad
\pi_m(\bz)=\frac{\exp(-\widetilde{\operatorname{dist}}_m(\bz)/\tau)}{\sum_k \exp(-\widetilde{\operatorname{dist}}_k(\bz)/\tau)},
\end{equation*}
with $\tau = 1$. $A_m$ is not normalised across models, so one patient can score high on several of them. Clinical metadata were excluded from model training and analysed only after patient embeddings had been frozen.

\paragraph{Drug treatments} Doses, vehicles, routes and timing for each model--drug arm. Logged timing is the median [range] interval in minutes between drug administration and the convulsant.

\begin{center}
\footnotesize
\setlength{\tabcolsep}{3pt}
\begin{tabular}{L{0.88in}L{0.52in}L{1.12in}L{0.72in}L{0.48in}L{1.12in}L{0.42in}}
\toprule
\textbf{Drug (vial)} & \textbf{Dose} & \textbf{Vehicle} & \textbf{Route} & \textbf{Pre-treat} & \textbf{Logged timing (min)} & \textbf{\emph{n} (M/F)} \\
\midrule
\multicolumn{7}{l}{\textit{4-AP, 7.5 mg/kg i.p.}} \\
Soticlestat (10 mg) & 10 mg/kg & saline & PO gavage (10 $\mu$L/g) & 60 min & 60.3 [59.2--60.5] & 8 (5/3) \\
Padsevonil (1 mg) & 3 mg/kg & 20\% HP-$\beta$-CD & i.p. & 30 min & 30.3 [30.2--30.7] & 8 (5/3) \\
Ganaxolone & 10 mg/kg & 30\% HP-$\beta$-CD in saline & s.c. & 15 min & 15.0 [14.8--15.5] & 12 (6/6) \\
JNJ-40411813 & 20 mg/kg & 20\% HP-$\beta$-CD + 1 eq HCl, pH 4 & s.c. & 30 min & 30.3 [30.0--31.0] & 11 (5/6) \\
Valproate & 400 mg/kg & 20\% sucrose & PO gavage (8 $\mu$L/g) & 30 min & 30.5 [30.1--31.1] & 10 (5/5) \\
\addlinespace[2pt]
\multicolumn{7}{l}{\textit{PTZ, 50 mg/kg i.p.}} \\
Tiagabine$\cdot$HCl (10 mg) & 10 mg/kg & saline & i.p. & 30 min & 30.3 [30.2--30.3] & 8 (5/3) \\
Levetiracetam (100 mg) & 100 mg/kg & saline & i.p. & 60 min & 60.2 [59.8--60.3] & 8 (5/3) \\
\addlinespace[2pt]
\multicolumn{7}{l}{\textit{AY9944, 7.5 mg/kg s.c. (P2, P8, P14, P20)}} \\
Tiagabine$\cdot$HCl & 10 mg/kg & saline & i.p. & n/a, record from injection & Drug at $t$ = 4 h; 120.3 [120.2--122.5] after vehicle & 8 (4/4) \\
Ethosuximide & 250 mg/kg & saline & i.p. & n/a, record from injection & Drug at $t$ = 4 h; 120.3 [120.0--120.7] after vehicle & 16 (8/8) \\
Valproate & 400 mg/kg & 20\% sucrose & PO gavage (8 $\mu$L/g) & n/a, record from injection & Drug at $t$ = 4 h; 121.5 [120.0--123.4] after vehicle & 16 (8/8) \\
\bottomrule
\end{tabular}
\end{center}

\paragraph{Pharmacological perturbation analysis} Treated mouse recordings were entirely excluded from representation learning, checkpoint selection and definition of the disease manifolds. The trained model was frozen before treated recordings were embedded.

For AY9944 and PTZ, pre- and post-treatment recordings were available from the same animal, allowing response to be expressed as displacement from the individual disease state towards the healthy representation. In 4-AP, treatment preceded disease induction; treated animals were therefore evaluated relative to the independently defined untreated 4-AP disease manifold and vehicle group.

Treatment displacement was decomposed into movement parallel to the disease-to-healthy trajectory, used as the neural-rescue metric, and orthogonal displacement, analysed exploratorily as a measure of additional pharmacological effects. Writing $\Delta=\bz_{\text{treated}}-\bz_{\text{anchor}}$, with the anchor the animal's own pre-dose or vehicle epoch moved onto the target $T$ ($T=C_m^{D}$ for AY9944 and PTZ, $T=C_H$ for 4-AP) to give $\bz_d$, the local route to healthy was built from the untreated human controls and model-$m$ controls $h_i \in \mathcal{H}$ with $\delta_i=\lVert h_i-\bz_d\rVert$:
\begin{equation*}
w_i \propto \exp\!\left(-\frac{\delta_i-\min_j \delta_j}{\tau\,\operatorname{sd}(\delta)}\right),
\qquad
r=\sum_i w_i h_i-\bz_d,
\qquad
M_2=\frac{\Delta\cdot r}{\lVert r\rVert^{2}} .
\end{equation*}
$M_2$ is the fraction of the route travelled: $1$ means the animal reached the healthy barycentre and a negative value means it moved away. Orthogonal displacement was measured outside a discriminative subspace $B\in\mathbb{R}^{4\times d}$ with orthonormal rows, where $b_1$ is the normalised Fisher direction $S_w^{-1}(\mu_D-\mu_H)$ and $b_2$ to $b_4$ are the leading eigenvectors by $|\lambda|$ of $L^{\mathsf T}(\Sigma_D-\Sigma_H)L$ with $LL^{\mathsf T}=S_w^{-1}$, mapped back through $L$ and Gram--Schmidt orthogonalised:
\begin{equation*}
\operatorname{orth}=\lVert (I-B^{\mathsf T}B)\,\Delta\rVert,
\qquad
\operatorname{orth}_{\text{frac}}=\frac{\operatorname{orth}}{\lVert \Delta\rVert},
\qquad
\operatorname{orth}_{\text{routes}}=\frac{\operatorname{orth}}{\lVert r\rVert} .
\end{equation*} Because the paired AY9944/PTZ experiments and 4-AP used different estimators, arm-level rescue scores were standardized within the experimental family before pooling.

Human efficacy rankings were defined independently before analysis of treated EEG. Their association with neural rescue was quantified using Spearman correlation. Significance was evaluated using a two-sided exact blocked permutation test within the experimental family (all 14,400 relabellings), and the pooled confidence interval was estimated from 4,000 bootstrap resamples of animals within treatment arms.

\paragraph{Clinical efficacy benchmark} Before any efficacy metric was computed, each of the ten model--drug combinations was assigned an ordinal human efficacy score from $-$1 to 4. Scores reflected the published clinical evidence for the seizure type the model best represents. The scale ran from 4 (regulatory approval and positive randomized controlled trials, first-line or broad-spectrum), through 3 (narrower or second-line efficacy) and 2 (approval only for another seizure type, as adjunctive therapy). Scores of 1 and 0 were given to compounds whose clinical trials did not meet their primary efficacy endpoints: 1 when some supportive clinical signal remained (JNJ-40411813, soticlestat), and 0 when the failed trial was followed by discontinued development (padsevonil). A score of $-$1 marked documented clinical aggravation of the seizure type (tiagabine in absence epilepsy). The resulting scores were:

\begin{center}
\footnotesize
\begin{tabular}{llc}
\toprule
\textbf{Drug} & \textbf{Model} & \textbf{Frozen human efficacy} \\
\midrule
Valproate      & 4-AP   & 4 \\
Ethosuximide   & AY9944 & 4 \\
Levetiracetam  & PTZ    & 4 \\
Ganaxolone     & 4-AP   & 3 \\
Valproate      & AY9944 & 3 \\
Tiagabine      & PTZ    & 2 \\
JNJ-40411813   & 4-AP   & 1 \\
Soticlestat    & 4-AP   & 1 \\
Padsevonil     & 4-AP   & 0 \\
Tiagabine      & AY9944 & $-$1 \\
\bottomrule
\end{tabular}
\end{center}

\paragraph{Off-target rubric} Before any latent metric was computed, each drug was scored from its human clinical profile (FDA labels and pivotal trials) on three 1--4 scales: CNS sedation, psychoactive effects and systemic (non-CNS) adverse events. Scores were generated by a large language model (Claude Opus 5, Anthropic) and checked by the authors. The composite off-target score was the mean of the CNS sedation and psychoactive scores, raised or lowered by 0.5 when systemic burden was above or below a typical level of 2. Scores were assigned per drug and applied to every mouse model in which the drug was tested.

\paragraph{Conventional EEG readouts} EEG from two channels was resampled to 250 Hz, band-pass filtered between 0.5 and 70 Hz (4th-order zero-phase Butterworth) and divided into 1-s epochs. For each epoch, RMS amplitude and line length (mean absolute first difference) were averaged across channels. Each was z-scored against the same animal's own pre-drug baseline. For 4-AP and PTZ this was the 30-min pre-drug recording; for AY9944, which are chronically epileptic, it was the animal's untreated recording. Normalizing within each animal removes between-animal differences in electrode gain. In the post-drug test epoch (30 min; 32 min for AY9944), an epoch was flagged when its RMS or line-length z-score was $\geq$ 4. Flagged epochs separated by $\leq$ 2 s were merged into events, and events shorter than 2 s were discarded. Seizure count was the number of events per animal, and time in seizure the summed event duration as a fraction of the epoch. Mean amplitude shift was the mean RMS z-score across the epoch, and high-amplitude time the fraction of epochs with an RMS z-score $\geq$ 4. Each readout was averaged per arm and signed so that less seizure activity scored higher. It was then z-scored within model family (AY9944/PTZ and 4-AP) and tested against human efficacy exactly as the cross-species biomarker was: pooled Spearman correlation with a two-sided exact blocked permutation test within the experimental family (all 14,400 relabellings).

\paragraph{\emph{Fmr1}-16p11.2 analysis} Human 16p11.2 EEG was acquired at 500 Hz using a 128-channel EGI HydroCel system. After removal of reference, absent and ocular channels, 117 candidate EEG channels remained. The cohort comprised 14 controls, 25 deletion carriers and 15 duplication carriers.

Mouse recordings comprised 32 Neuronexus contacts selected from 64-channel Open Ephys acquisitions recorded at 30 kHz from 53 wild-type and 17 \emph{Fmr1}-knockout mice. Mouse signals were progressively resampled to 250 Hz and human recordings from 500 to 250 Hz. Both datasets were processed within a common 1--95-Hz band and divided into 8-s windows. Normalization was performed independently within each window so that no statistic crossed subject boundaries.

Dual-rule learning aligned wild-type mice with human controls and \emph{Fmr1}-knockout mice with 16p11.2 carriers. Deletion and duplication carriers were combined during training but analysed separately afterwards. The alignment score projects an embedding onto the mouse disease direction, taken from the training units $\mathcal{T}$:
\begin{equation*}
\hat{u}_M=\frac{\mu(\mathcal{T}\cap\mathrm{KO})-C_{\mathrm{WT}}}{\lVert \mu(\mathcal{T}\cap\mathrm{KO})-C_{\mathrm{WT}}\rVert},
\qquad
a(\bz)=(\bz-C_{\mathrm{WT}})\cdot\hat{u}_M,
\qquad
a_z(\bz)=\frac{a(\bz)-\operatorname{mean}_{\mathcal{T}\cap\mathrm{human}} a}{\operatorname{sd}_{\mathcal{T}\cap\mathrm{human}} a},
\end{equation*}
where $C_{\mathrm{WT}}=\mu(\mathcal{T}\cap\mathrm{WT})$ and $a_z$ is the reported alignment score. Participant-level alignment was averaged across five independently trained model seeds.

\paragraph{Statistical analysis} The participant or animal was the unit of biological replication. Individual EEG windows were never treated as independent replicates. Classification performance was summarized using AUROC or balanced accuracy. Pearson correlation was used for spectral-profile similarity and Spearman correlation for ordinal translational associations. Non-parametric group (e.g. Kruskal-Wallis test) and paired comparisons were used where appropriate, with Bonferroni-Holm correction for families of multiple comparisons. The Kruskal-Wallis test was applied to the displacement of the six treatment groups from the untreated disease state in the 4-AP model, and to participant-level alignment scores across the three human genotype groups in the \emph{Fmr1}-16p11.2 analysis. Bootstrap and permutation analyses preserved the relevant animal, participant and experimental-family structure.

\paragraph{Ethics} Animal procedures were approved by the Institutional Animal Care and Use Committee of Charles River Accelerator and Development Lab in South San Francisco (protocol 2024-2460, ``Investigating brain activity from molecular mechanisms to systems-level dynamics under neuronal activity modulation'', approved 20th January 2025) and by the Institutional Animal Care and Use Committee of Exin Therapeutics Philippines Representative Office registered with the Bureau of Animal Industry (protocol 2025-EX-CS0109 approved in 18th December 2025). Procedures were performed in accordance with the Guide for the Care and Use of Laboratory Animals and the Philippine Animal Welfare Act (RA 8485, as amended by RA 10631).

Nine healthy adult volunteers (6 male, 3 female) underwent non-invasive scalp EEG during audiovisual stimulation. The protocol (2025-EX-HU0102) was approved by the internal research ethics committee of Exin Therapeutics on 3rd March 2026, and all participants gave written informed consent.

Human EEG from the Temple University Hospital EEG Corpus and the Simons Searchlight dataset (SFARI Base) was obtained under the respective data use agreements and was de-identified before access.

\section{Data availability}

Human EEG data analysed in this study were obtained from two third-party resources. The Temple University Hospital EEG Corpus, including the Temple University Hospital Seizure Corpus (TUSZ) and the Temple University Hospital Epilepsy Corpus (TUEP), is distributed by the Neural Engineering Data Consortium at Temple University and is available at \url{https://isip.piconepress.com/projects/nedc/html/tuh_eeg/}\cite{r27,r28}. The human 16p11.2 EEG recordings were obtained from the Simons Searchlight EEG Data resource on SFARI Base (\url{https://base.sfari.org/}) and are available to approved researchers through that portal. Mouse EEG and electrophysiological recordings generated in this study are available from the corresponding author on reasonable request. Code and rodent data will be released upon peer reviewed publication.

\section{Acknowledgements}

We are grateful to all of the families at the participating Simons Searchlight sites as well as the Simons Searchlight Consortium, formerly the Simons VIP Consortium. We appreciate obtaining access to electroencephalography (EEG) data on SFARI Base. Approved researchers can obtain the Simons Searchlight population dataset described in this study by applying at \url{https://base.sfari.org/}.

We also acknowledge the Neural Engineering Data Consortium at Temple University for making the Temple University Hospital EEG Corpus publicly available.

\section{Author contributions}

G.O.-S. and M.T. conceived the study and designed the experiments. G.O.-S. developed the computational framework and performed the analyses. All authors contributed to data collection and experimental validation. G.O.-S. wrote the manuscript and M.T. reviewed and edited it. G.O.-S. supervised the project. All authors read and approved the final manuscript. Large language models were used to assist with the writing and editing of this manuscript; the authors reviewed and verified all content and take full responsibility for it.

\section{Competing interests}

All authors are employees of Exin Therapeutics, Inc. M.T. and G.O.-S. hold equity in Exin Therapeutics, Inc. A provisional patent application covering the cross-species neural representation learning methods described in this manuscript has been filed by Exin Therapeutics, Inc.

\renewcommand{\refname}{Bibliography}
\begingroup
\small
\setlength{\parskip}{0pt}

\endgroup

\ifdefined\suppstandalone\else
\clearpage
\section{Supplemental}
\fi
\setcounter{table}{0}
\captionsetup[table]{name=Supplementary Table,labelsep=period,skip=6pt,font=footnotesize}

\begin{table}[!ht]
\centering
\footnotesize
\setlength{\tabcolsep}{4.5pt}
\renewcommand{\arraystretch}{1.05}
\begin{tabular}{lccccccc}
\toprule
\textbf{Task} &
\begin{tabular}[b]{@{}c@{}}\textbf{Subjects}\\\textbf{tested}\end{tabular} &
\begin{tabular}[b]{@{}c@{}}\textbf{Projection}\\\textbf{axis}\end{tabular} &
\begin{tabular}[b]{@{}c@{}}\textbf{Mean}\\\textbf{projection}\end{tabular} &
\textbf{Sign} &
\begin{tabular}[b]{@{}c@{}}\textbf{Sample}\\\textbf{size}\end{tabular} &
\textbf{\emph{P}-value} &
\begin{tabular}[b]{@{}c@{}}\textbf{FDR-adjusted}\\\textbf{\emph{P} (BH)}\end{tabular} \\
\midrule
\textbf{Audio vs Visual}  & Human & Human & $\approx$0.604   & $+$ & 9  & 0.03516    & \textbf{0.04280}  \\
\textbf{Audio vs Visual}  & Human & Mouse & $\approx$0.606   & $+$ & 9  & 0.03516    & \textbf{0.04280}  \\
\textbf{Audio vs Visual}  & Mouse & Mouse & $\approx$0.815   & $+$ & 12 & 0.0004883  & \textbf{0.002279} \\
\textbf{Audio vs Visual}  & Mouse & Human & $\approx$0.821   & $+$ & 12 & 0.0004883  & \textbf{0.002279} \\
\addlinespace[2pt]
\textbf{ASSR}             & Human & Human & $\approx$0.703   & $+$ & 9  & 0.01953    & \textbf{0.02734}  \\
\textbf{ASSR}             & Human & Mouse & $\approx$0.658   & $+$ & 9  & 0.01953    & \textbf{0.02734}  \\
\textbf{ASSR}             & Mouse & Mouse & $\approx$0.651   & $+$ & 12 & 0.006836   & \textbf{0.01562}  \\
\textbf{ASSR}             & Mouse & Human & $\approx$0.642   & $+$ & 12 & 0.007324   & \textbf{0.01562}  \\
\addlinespace[2pt]
\textbf{Chirp}            & Human & Human & $\approx$0.578   & $+$ & 9  & 0.02734    & \textbf{0.03645}  \\
\textbf{Chirp}            & Human & Mouse & $\approx$0.560   & $+$ & 9  & 0.04297    & 0.05013           \\
\textbf{Chirp}            & Mouse & Mouse & $\approx$0.535   & $+$ & 12 & 0.04736    & 0.05304           \\
\textbf{Chirp}            & Mouse & Human & $\approx$0.581   & $+$ & 12 & 0.01953    & \textbf{0.02734}  \\
\addlinespace[2pt]
\textbf{Oddball SV--DV}   & Human & Human & $\approx-$0.471  & $-$ & 9  & 0.07422    & 0.07993           \\
\textbf{Oddball SV--DV}   & Human & Mouse & $\approx$0.111   & $+$ & 9  & 0.8633     & 0.8633            \\
\textbf{Oddball SV--DV}   & Mouse & Mouse & $\approx-$0.589  & $-$ & 12 & 0.01514    & \textbf{0.02573}  \\
\textbf{Oddball SV--DV}   & Mouse & Human & $\approx$0.163   & $+$ & 12 & 0.7324     & 0.7595            \\
\addlinespace[2pt]
\textbf{Checkerboard}     & Human & Human & $\approx$0.662   & $+$ & 9  & 0.01562    & \textbf{0.02573}  \\
\textbf{Checkerboard}     & Human & Mouse & $\approx$0.660   & $+$ & 9  & 0.01562    & \textbf{0.02573}  \\
\textbf{Checkerboard}     & Mouse & Mouse & $\approx$0.814   & $+$ & 12 & 0.0009766  & \textbf{0.003418} \\
\textbf{Checkerboard}     & Mouse & Human & $\approx$0.819   & $+$ & 12 & 0.0009766  & \textbf{0.003418} \\
\addlinespace[2pt]
\textbf{Gratings}         & Human & Human & $\approx$0.755   & $+$ & 9  & 0.003906   & \textbf{0.01094}  \\
\textbf{Gratings}         & Human & Mouse & $\approx$0.778   & $+$ & 9  & 0.003906   & \textbf{0.01094}  \\
\textbf{Gratings}         & Mouse & Mouse & $\approx$0.973   & $+$ & 12 & 0.0004883  & \textbf{0.002279} \\
\textbf{Gratings}         & Mouse & Human & $\approx$0.973   & $+$ & 12 & 0.0004883  & \textbf{0.002279} \\
\addlinespace[2pt]
\textbf{Optic flow}       & Human & Human & $\approx$0.758   & $+$ & 9  & 0.007812   & \textbf{0.01562}  \\
\textbf{Optic flow}       & Human & Mouse & $\approx$0.726   & $+$ & 9  & 0.007812   & \textbf{0.01562}  \\
\textbf{Optic flow}       & Mouse & Mouse & $\approx$0.939   & $+$ & 12 & 0.0004883  & \textbf{0.002279} \\
\textbf{Optic flow}       & Mouse & Human & $\approx$0.937   & $+$ & 12 & 0.0004883  & \textbf{0.002279} \\
\bottomrule
\end{tabular}
\caption{\textbf{Subject-level validation of within- and cross-species sensory-response axes.} Mean subject-level projections onto species-derived sensory-response axes for each sensory paradigm and for the pooled analysis. For each task, responses from human or mouse subjects were projected onto either the axis derived from the same species (own axis) or the axis independently derived from the other species (cross-species axis). Positive projection indicates that the subject-level stimulus response is oriented in the predicted direction of the corresponding sensory-response axis. The table reports the mean projection, projection sign, number of subjects ($n$), raw \emph{P}-value and Benjamini-Hochberg false-discovery-rate-adjusted \emph{P}-value. Multiple-comparison correction was applied across all 28 reported subject-level tests. Human $n$ = 9; mouse $n$ = 12. Bold adjusted \emph{P}-values indicate FDR $<$ 0.05}
\label{tab:1}
\end{table}

\begin{table}[htbp]
\centering
\footnotesize
\setlength{\tabcolsep}{4pt}
\renewcommand{\arraystretch}{1.1}
\begin{tabular}{L{1.30in}L{1.52in}L{1.52in}L{1.52in}}
\toprule
\textbf{Attribute} & \textbf{Mouse PTZ} & \textbf{Mouse 4-AP} & \textbf{Mouse AY9944} \\
\midrule
\textbf{Mouse contrast} & Discharge vs duration-matched non-seizure (within animal) & Detected ictal vs non-seizure (within animal) & SWD vs the same detector's positives in saline animals (between animal) \\
\addlinespace[2pt]
\textbf{Human comparator} & TUSZ, by seizure type vs healthy & TUSZ, by seizure type vs healthy & TUSZ, by seizure type vs healthy \\
\addlinespace[2pt]
\textbf{Mouse N} & 13 animals / 57 sessions (Control arm) & 23 animals & 32 untreated + 32 saline animals / 233 recordings \\
\addlinespace[2pt]
\textbf{Spectral peak (Hz)} & 3 & 29 & 3 \\
\addlinespace[2pt]
\textbf{Gamma 30-70 Hz enhancement} & -20.4 & +22.2 & -4.0 (suppressed) \\
\addlinespace[2pt]
\textbf{Best human match} & ABSZ +0.94 & TCSZ +0.64 & ABSZ +0.91 \\
\addlinespace[2pt]
\textbf{Next best} & FNSZ +0.84 $\cdot$ CPSZ +0.68 & TNSZ +0.36 $\cdot$ GNSZ -0.20 & FNSZ +0.86 $\cdot$ GNSZ +0.72 \\
\addlinespace[2pt]
\textbf{Worst human match} & TCSZ -0.30 $\cdot$ TNSZ -0.29 & FNSZ -0.58 $\cdot$ CPSZ -0.57 & TNSZ -0.30 $\cdot$ TCSZ +0.05 \\
\addlinespace[2pt]
\textbf{Transfer m-$>$h} & 0.89 & 0.81 & 0.86 \\
\addlinespace[2pt]
\textbf{Transfer h-$>$m} & 0.75 & 0.55 & 0.55 \\
\bottomrule
\end{tabular}
\caption{\textbf{Summary of cross-species EEG alignment across mouse seizure models and human epilepsy phenotypes.} Comparison of mouse contrast definitions, human reference cohorts, sample sizes, spectral characteristics, human seizure-type matches, bidirectional transfer performance, and seizure-type specificity across PTZ, 4-AP, and AY9944 analyses.}
\label{tab:2}
\end{table}

\clearpage

\begingroup
\footnotesize
\setlength{\tabcolsep}{4pt}
\renewcommand{\arraystretch}{1.12}
\begin{longtable}{L{1.12in}L{2.38in}L{2.38in}}
\caption{\textbf{Training and evaluation configuration for the two cross-species models.} Dataset composition, preprocessing, architecture, loss terms with their weights, optimization, cross-validation and selection, and evaluation, for the unified epilepsy model and for the \emph{Fmr1}--16p11.2 model.}
\label{tab:3}\\
\toprule
 & \textbf{Unified epilepsy model} & \textbf{\emph{Fmr1}--16p11.2 model} \\
\midrule
\endfirsthead
\multicolumn{3}{l}{\emph{Supplementary Table \thetable, continued}}\\
\toprule
 & \textbf{Unified epilepsy model} & \textbf{\emph{Fmr1}--16p11.2 model} \\
\midrule
\endhead
\midrule
\multicolumn{3}{r}{\emph{continued on the next page}}\\
\endfoot
\bottomrule
\addlinespace[3pt]
\multicolumn{3}{@{}p{5.95in}@{}}{\footnotesize $^{1}$ A channel is rejected if the robust $z$ of its log SD or of its line-noise ratio exceeds 4, or if its maximum correlation with any other channel is below 0.25. If more than half the channels would be rejected, all are kept.\par $^{2}$ Outlying channel: robust $z$ of log peak-to-peak amplitude $>$ 6.\par $^{3}$ Composite = mean(human AUROC, mouse AUROC) $-$ 0.5$\cdot$min(cross-species MMD$^{2}$ + CORAL, 0.5) + 0.05$\cdot$min(prototype separation, 2) + 0.2$\cdot$(fraction of validation epilepsy patients on the disease side of the healthy--disease midpoint). It is computed on validation subjects only.}\\
\endlastfoot

\multicolumn{3}{@{}l}{\textbf{Data}}\\
\addlinespace[2pt]
Human units & TUEP: 96 patients with epilepsy, 82 without; each patient's recordings ($\leq$10) pooled into one unit & 14 controls, 25 deletion and 15 duplication carriers; carriers pooled as one class in training \\
Mouse units & PTZ, 13 mice: 0--30 min post-PTZ vs 0--30 min post-saline, same animal\newline 4-AP, 14 mice: 0--60 min post-4-AP vs pre-injection epoch, same animal\newline AY9944: 43 untreated or control-vector mice vs 32 saline controls & 53 wild-type, 17 \emph{Fmr1}-KO \\
Channels & Mouse: EEG1, EEG2\newline Human: 2 drawn at random per window from 19 referential 10--20 channels & 117 human and 32 mouse candidates; 97--117 and 25--32 remain after bad-channel rejection$^{1}$; 16 drawn at random per window in both species \\
Resampling and filtering & 250 Hz; 0.5--70 Hz band-pass (4th-order zero-phase Butterworth); no notch & Mouse: 30 kHz $\rightarrow$ 1.5 kHz (FIR), notch at 60 Hz and harmonics, 1--95 Hz band-pass, $\rightarrow$ 250 Hz\newline Human: notches at 60--240 Hz, 1--95 Hz band-pass, 500 $\rightarrow$ 250 Hz\newline Both: 56--64 Hz band-stop; all filters zero-phase, 4th-order Butterworth \\
Windows & 8 s (2,000 samples), evenly spaced; $\leq$400 per human recording, $\leq$240 per mouse epoch & 8 s, $\sim$240 per recording (step 2--8 s); a window is rejected if $>$25\% of its channels are outliers$^{2}$ \\
Normalization & Per window and channel: ($x-$median)/IQR & The same, plus a learnable gain and offset per species \\
Windows per unit per step & 64 (drawn with replacement if fewer are available) & 16 (without replacement) \\
Augmentation & Time reversal and polarity flip, $p$ = 0.5 each & Sign flip and channel permutation, $p$ = 0.5 each \\
\addlinespace[4pt]

\multicolumn{3}{@{}l}{\textbf{Architecture}}\\
\addlinespace[2pt]
Species-specific input & Per-channel affine + 2$\times$2 channel mixing, identity-initialized (20 parameters) & Gain + offset per species (4 parameters) \\
Filterbank (shared) & Learnable sinc: 32 band-pass filters, 129 taps, Hamming window, log-spaced initialization & Same \\
Temporal stem (shared) & Depthwise conv (k15, s4) $\rightarrow$ 1$\times$1 conv 32$\rightarrow$64 $\rightarrow$ conv 64$\rightarrow$64 (k15, s4); BN + GELU after each; averaged over the 2 channels & Same, applied per channel \\
Encoder (shared) & Conv 64$\rightarrow$128 (k7, s2) $\rightarrow$ 128$\rightarrow$128 (k5, s2) $\rightarrow$ 128$\rightarrow$128 (k3); BN + GELU; global average pool; dropout 0.3; linear $\rightarrow$ 64-D window embedding & Same \\
Channel pooling & Mean (after the stem) & Attention (64$\rightarrow$32$\rightarrow$1, tanh) + mean, then MLP 128$\rightarrow$64$\rightarrow$64 \\
Window pooling & DeepSets: $\varphi$ 64$\rightarrow$96$\rightarrow$96, [mean $\|$ max], $\rho$ 192$\rightarrow$96 & Mean over 16 windows, then MLP 64$\rightarrow$64$\rightarrow$48 \\
Latents & $z_{\mathrm{trans}}$ 14-D and $z_{\mathrm{private}}$ 16-D (heads 96$\rightarrow$96$\rightarrow d$) & Shared 32-D and private 16-D (BatchNorm; shared then LayerNorm) \\
Output heads & Linear state head on BN($z_{\mathrm{trans}}$); linear species head on BN($z_{\mathrm{private}}$) & Linear disease head on the shared latent; species MLP (16$\rightarrow$32$\rightarrow$1) on the private latent \\
Parameters & 318,107 & 286,008 \\
\addlinespace[4pt]

\multicolumn{3}{@{}l}{\textbf{Loss (weight)}}\\
\addlinespace[2pt]
Supervised contrastive & Within each state: $\tau$ = 0.2; batch-centred, L2-normalized; positives are cross-species pairs only (included in the two alignment terms below) & $\tau$ = 0.2; positives are same-state pairs; cross-species healthy pairs $\times$4; cross-species disease pairs neutral (removed from numerator and denominator); already-aligned positives (cosine $\geq$ 0.4) excluded from the contrastive term (1) \\
Healthy alignment & Contrastive + MMD$^{2}$ + CORAL, human vs pooled mouse (2) & MMD$^{2}$ + CORAL + standardized centroid distance, control vs WT (3) \\
Disease alignment & Contrastive + MMD$^{2}$ + CORAL, human vs pooled mouse (2) & MMD$^{2}$, CNV vs KO (0.5) \\
MMD and CORAL details & RBF kernel, bandwidths \{0.25, 0.5, 1, 2, 4\} $\times$ median squared distance; CORAL = $\lVert \mathrm{Cov}_{a} - \mathrm{Cov}_{b} \rVert_{F}^{2} / 4d^{2}$ & Same \\
Disease-state classifier & Binary cross-entropy (6) & Binary cross-entropy (0.5) \\
Model preference & Soft minimum (temperature 1) of each patient's scaled distance to the 3 mouse disease prototypes (0.15) & -- \\
Prototype separation & Hinge, margin 1 (0.3) & -- \\
Species on private latent & Binary cross-entropy (1) & Binary cross-entropy (1) \\
Shared--private independence & RBF HSIC + cross-covariance (2) & RBF HSIC (1) \\
Anti-collapse & SD $\geq$ 0.35 per dimension (1); off-diagonal covariance (1); relu(1 $-$ PR/10)$^{2}$ (4) & SD $\geq$ 0.35 (1); covariance (1); relu(1 $-$ PR/6)$^{2}$ (3) \\
Diagnostic species probe & -- & Binary cross-entropy on the detached shared latent (1); passes no gradient to the encoder \\
\addlinespace[4pt]

\multicolumn{3}{@{}l}{\textbf{Optimization}}\\
\addlinespace[2pt]
Optimizer & AdamW: learning rate $3\times10^{-4}$, weight decay 0.01, $\beta$ 0.9/0.999, $\varepsilon$ $10^{-8}$ & Same \\
Schedule & 2-epoch linear warm-up, then cosine decay to 0, stepped per batch & Same \\
Clipping and precision & Gradient norm $\leq$ 5; fp32 & Same \\
Batch & 24 units (3 from each of 8 species $\times$ model $\times$ state groups) $\times$ 64 windows $\times$ 2 channels & 16 units (4 from each of 4 groups) $\times$ 16 windows $\times$ 16 channels \\
Training length & 80 batches $\times$ 16 epochs = 1,280 steps; no early stopping & 64 batches $\times$ 25 epochs = 1,600 steps; no early stopping \\
Hardware & 1 NVIDIA L4, $\sim$6.5 min per run & 1 NVIDIA A10G, $\sim$5 min per run \\
\addlinespace[4pt]

\multicolumn{3}{@{}l}{\textbf{Cross-validation and selection}}\\
\addlinespace[2pt]
Folds & 3 fixed subject-level folds, stratified by group & 5 subject-level folds, stratified by group, redrawn for each seed \\
Models & 4 seeds $\times$ 3 folds = 12 & 5 seeds $\times$ 5 folds = 25 \\
Validation set & 20\% of training subjects per group & 20\% of training subjects per group \\
Checkpoint & Epoch maximizing a validation-only composite$^{3}$, among epochs with validation PR $\geq$ 4 & Final epoch; no checkpoint selection because the per-fold validation set was too small for stable model selection \\
Hyperparameter choice & Best of 6 configurations on the validation composite (fold 0, seed 0) & Channels per window $\in$ \{4, 8, 16, 32\} compared on validation; no difference, 16 kept \\
Excluded from training and selection & Test subjects; all drug-treated recordings & Test subjects \\
\addlinespace[4pt]

\multicolumn{3}{@{}l}{\textbf{Evaluation}}\\
\addlinespace[2pt]
Disease score & Projection onto the healthy $\rightarrow$ disease centroid axis from training units & Projection onto ($C_{\mathrm{disease}} - C_{\mathrm{healthy}}$), centroids averaged over species, z-scored within species on training subjects \\
Embedding draws per unit & 30 draws $\times$ 64 windows & 8 draws $\times$ 16 windows $\times$ 16 channels \\
Held-out AUROC & Subject scores averaged over seeds; 95\% CI from 2,000 class-stratified subject bootstraps & Same \\
Balanced accuracy & Youden-optimal threshold on training scores & Same \\
\end{longtable}
\endgroup

\clearpage

\setcounter{figure}{0}
\captionsetup[figure]{name=Supplementary Figure,labelsep=period}

\begin{figure}[htbp]
\centering
\includegraphics[width=\textwidth]{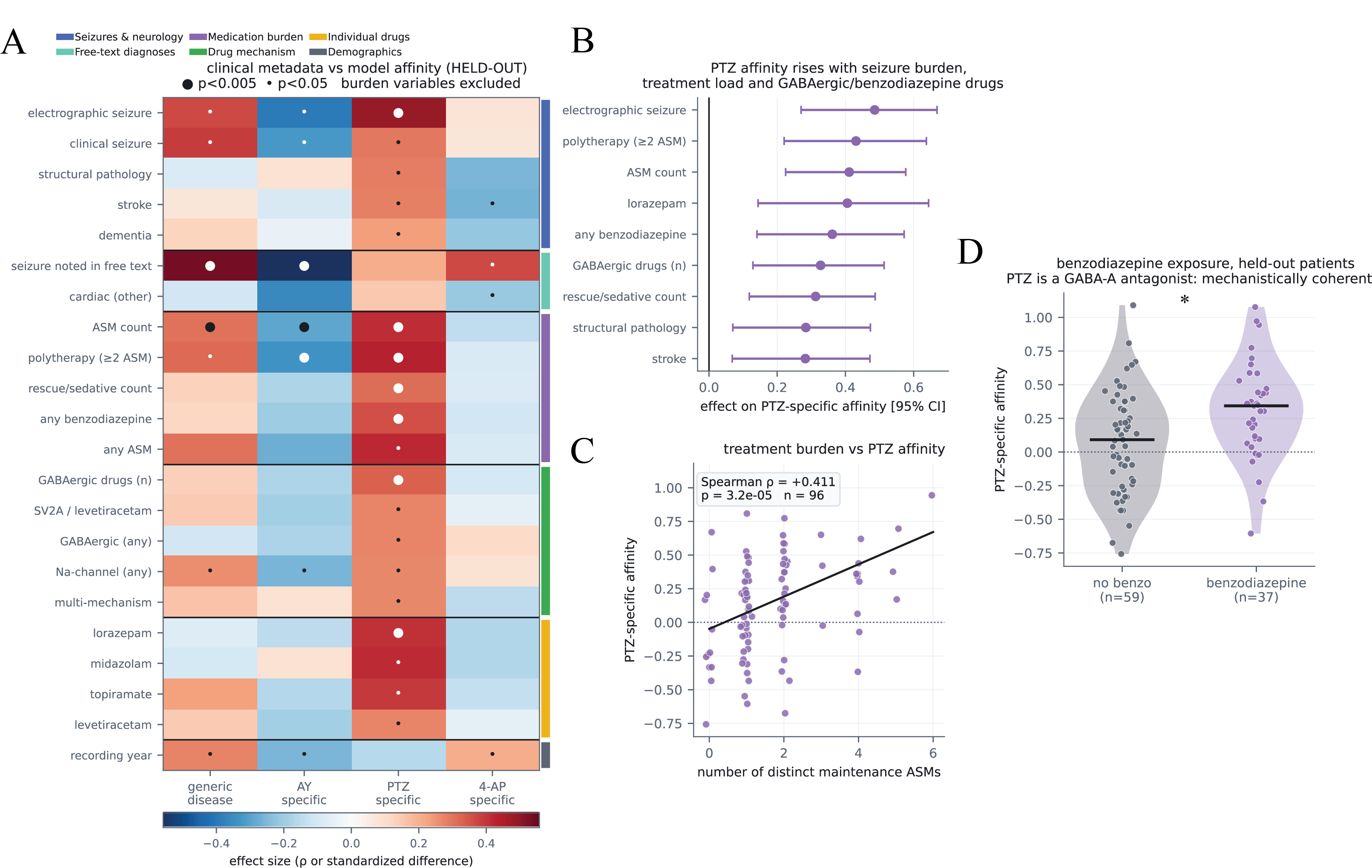}
\caption{\textbf{Clinical features associated with patient-specific alignment to mouse epilepsy models.} A, Association between clinical metadata and each patient's position in the cross-species latent space, quantified as affinity to a generic mouse epilepsy signature or to the AY9944, PTZ and 4-AP models. Colours indicate standardized effect size; points denote nominal significance ($P$ $<$ 0.05; large points, $P$ $<$ 0.005). Variables directly reflecting overall treatment burden were excluded from the generic disease analysis where indicated. B, PTZ-specific affinity was positively associated with multiple measures of seizure and antiseizure-medication burden, with particularly consistent associations for benzodiazepine and GABAergic drug exposure. Points and horizontal bars show effect estimates and uncertainty for individual clinical variables. C, Across patients, PTZ-specific affinity increased with the number of distinct maintenance antiseizure medications (Spearman $\rho$ = 0.411, $P$ = 3.2 $\times$ 10$^{-5}$, $n$ = 96). D, In held-out patients, benzodiazepine exposure was associated with greater PTZ-specific affinity (no benzodiazepine, $n$ = 59; benzodiazepine, $n$ = 37), consistent with the opposing modulation of GABA$_{\mathrm{A}}$ receptor signaling by benzodiazepines and the PTZ seizure model. These analyses indicate that individual patient--model alignment captures clinically and pharmacologically meaningful heterogeneity rather than a uniform epilepsy axis.}
\label{fig:S1}
\end{figure}

\clearpage

\begin{figure}[htbp]
\centering
\includegraphics[width=\textwidth]{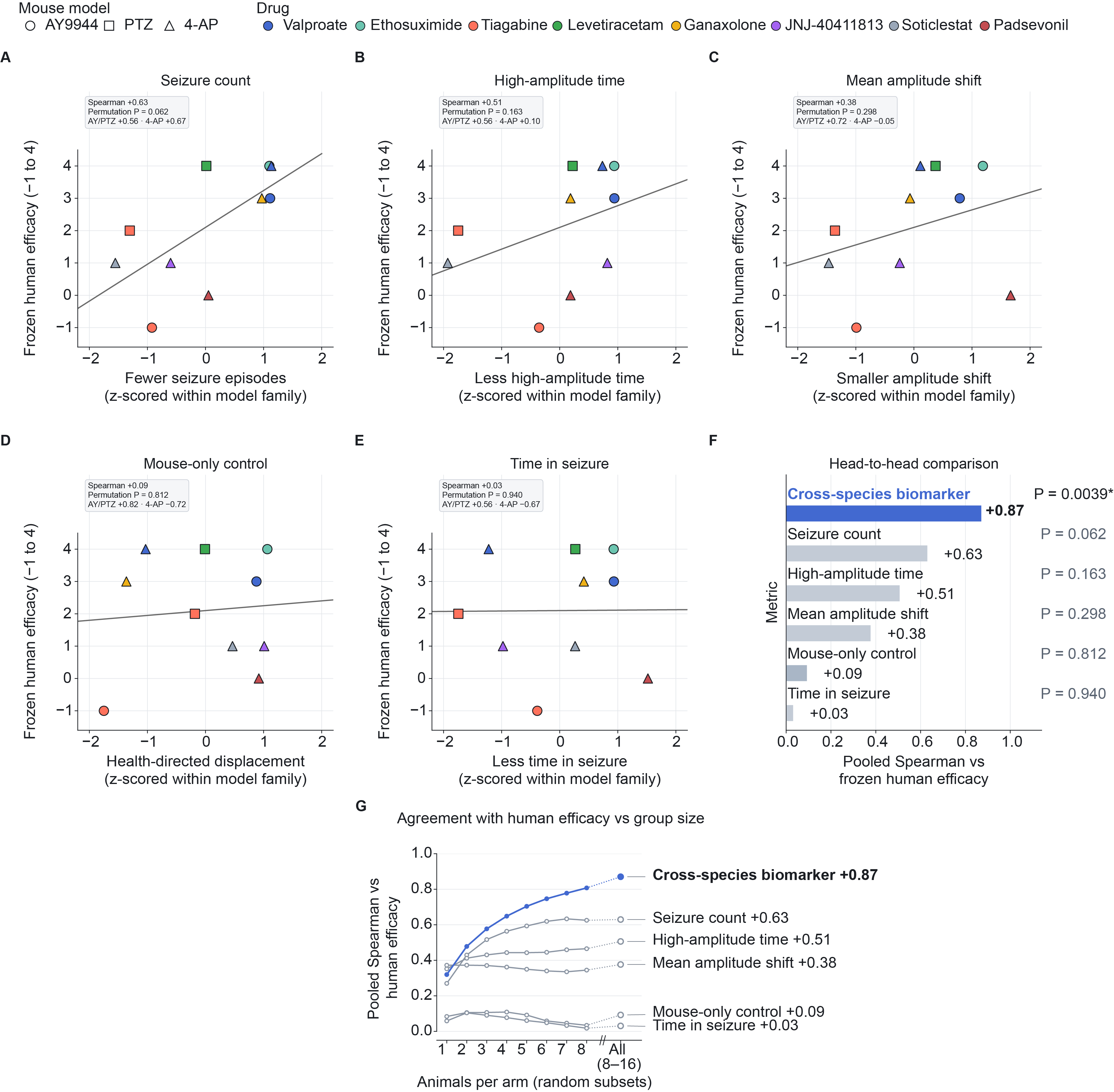}
\caption{\textbf{Conventional EEG seizure readouts track human clinical efficacy worse than the cross-species biomarker.} (A--E) Each panel plots one mouse EEG readout against frozen human efficacy (ordinal, $-$1 to 4) for the same 10 drug--model arms: 5 in 4-AP (triangles), 3 in AY9944 (circles) and 2 in PTZ (squares). Colours mark the drug, as in Figure~\ref{fig:4}. Each readout was averaged per arm over treated animals and z-scored within model family (4-AP; AY9944 + PTZ). Signs are set so that higher values always mean greater predicted rescue. (A) Number of seizure episodes per mouse in the 30 min after the convulsant. Episodes were detected algorithmically as $\geq$4$\sigma$ above the animal's own pre-drug baseline, lasting $\geq$2 s. (B) Fraction of 1 s epochs exceeding 4$\sigma$ of the animal's own baseline. (C) Mean EEG amplitude shift relative to the animal's own pre-drug baseline. (D) Mouse-only control: health-directed displacement computed from mouse recordings alone, with no human data. (E) Percentage of the 30 min epoch spent in detected seizures. Grey lines are least-squares fits, shown only as a guide. Insets give the pooled Spearman correlation and $P$ from a two-sided exact blocked permutation test within the experimental family (all 14,400 relabellings). They also give the within-family correlations. (F) Pooled Spearman correlation of each readout with human efficacy under this identical test, with the cross-species biomarker from Figure~\ref{fig:4} shown for reference (blue). *$P$ $<$ 0.05. None of the EEG readouts reached significance. The best, seizure count, reached $\rho$ = +0.63 ($P$ = 0.062). The cross-species biomarker reached $\rho$ = +0.87 ($P$ = 0.0039). (G) Pooled Spearman correlation with human efficacy as a function of the number of animals per arm, computed on random subsets of animals within each arm. Agreement with human efficacy increases with group size for the cross-species biomarker but not for the conventional readouts.}
\label{fig:S2}
\end{figure}

\end{document}